\documentclass[aps,prd,twocolumn, notitlepage,superscriptaddress,10pt,nofootinbib, 
amsmath,amssymb]{revtex4-1}

\usepackage{color}
\usepackage{graphicx} 
\usepackage{comment}

\newcommand{\fett}[1]{\boldsymbol{#1}}

\newcommand{\dd}{{\rm{d}}}

\newcommand{\ii}{{\rm{i}}}

\newcommand{\be}{\begin{equation}}
\newcommand{\ee}{\end{equation}}

\newcommand{\nabq}{\fett{\nabla}_{\fett{q}}}
\newcommand{\nabx}{\fett{\nabla}_{\fett{x}}}
\newcommand{\nab}{\fett{\nabla}}
\newcommand{\e}{{\rm e}}

\newcommand{\myArxiv}[1]{{\tt \href{http://arxiv.org/abs/#1}{{\color{black}[{\color{blue}#1}]}}}}

\newcommand{\DOI}[2]{ 
 \href{http://dx.doi.org/#1}{\color{blue} #2\!\!}
}

\definecolor{darkred}{rgb}{0.5,0,0}
\definecolor{darkgreen}{rgb}{0,0.5,0}
\definecolor{darkblue}{rgb}{0,0,0.5}

\newcommand\tildepi{\stackrel{\sim}{\smash{\pi}\rule{0pt}{0.7ex}}}

\usepackage{hyperref}
\hypersetup{colorlinks,
linkcolor=darkblue,
filecolor=darkgreen,
urlcolor=darkred,
citecolor=darkblue }

\begin{document}

\begin{flushright}
{
\tt 
  AEI-2014-043
}
\end{flushright}

\title{Relativistic Lagrangian displacement field and tensor perturbations}

\date{\today}

\author{Cornelius Rampf}
\email{cornelius.rampf@port.ac.uk}
\affiliation{Institute of Cosmology and Gravitation, University of Portsmouth, Portsmouth PO1 3FX, United Kingdom}
\affiliation{Max-Planck-Institute for Gravitational Physics (Albert-Einstein-Institute), D-14476 Potsdam-Golm, Germany}

\author{Alexander Wiegand}
\email{alexander.wiegand@aei.mpg.de}
\affiliation{Max-Planck-Institute for Gravitational Physics (Albert-Einstein-Institute), D-14476 Potsdam-Golm, Germany}

\begin{abstract}
We investigate the purely spatial Lagrangian coordinate transformation from the Lagrangian to the basic Eulerian frame. We demonstrate three techniques for extracting the relativistic displacement field from a given solution in the Lagrangian frame. These techniques are (a) from defining a local set of Eulerian coordinates embedded into the Lagrangian frame; (b) from performing a specific gauge transformation; and (c) from a fully non-perturbative approach based on the ADM split. The latter approach shows that this decomposition is not tied to a specific perturbative formulation for the solution of the Einstein equations. Rather, it can be defined at the level of the non-perturbative coordinate change from the Lagrangian to the Eulerian description.
Studying such different techniques is useful because it allows us to 
compare and develop further the various approximation techniques available in the Lagrangian formulation.
We find that one has to solve the gravitational wave equation in the relativistic analysis, otherwise the corresponding Newtonian limit will necessarily contain spurious non-propagating tensor artefacts at second order in the Eulerian frame. 
We also derive the magnetic part of the Weyl tensor in the Lagrangian frame, and find that it is not only excited by gravitational waves but also by tensor perturbations which are induced through the non-linear frame-dragging.  
We apply our findings to calculate for the first time the relativistic displacement field, up to second order, for a $\Lambda$CDM Universe in the presence of a local primordial non-Gaussian component.
Finally, we also comment on recent claims about whether mass conservation in the Lagrangian frame is violated.
\end{abstract}

\maketitle   

\section{Introduction}

Newtonian perturbation theory has been quite successful in describing the (mildly) non-linear regime of cosmological structure formation.
Its basic idea is to describe the cold dark matter (CDM) distribution of the Universe as an irrotational and pressureless fluid; inside the Newtonian regime the fluid evolution is governed by the Euler-Poisson system.
 Perhaps the most well-known approach is dubbed (Newtonian) Eulerian perturbation theory (NEPT) \cite{Bernardeau:2001qr}, where the Euler-Poisson system 
is solved with a perturbation Ansatz for the density and velocity fields. An alternative way to solve the Euler-Poisson system is to transform it to Lagrangian space, where the observer follows the gravitationally induced displacement of a given fluid element. This approach has only a single ``perturbation parameter'' which is the said displacement (field), and the approach is called Lagrangian perturbation theory (NLPT)  \cite{Zeldovich:1969sb,Buchert:1989xx,Bouchet:1994xp,Ehlers:1996wg,Rampf:2012xa,Rampf:2012up,Zheligovsky:2013eca,Villa:2014aja}. Depending on the specific application, either the Eulerian or Lagrangian picture could be favourable, although the Lagrangian approach contains always more non-linear information, and the Lagrangian series is expected to have better convergence properties, i.e., the Lagrangian solution remains significantly longer time-analytic as compared to the Eulerian solution \cite{Zheligovsky:2013eca}.\footnote{That only NLPT breaks down at shell-crossing but not NEPT is an unfortunate and common misunderstanding. Both NLPT and NEPT are fluid descriptions based on the Euler-Poisson system, so both NLPT and NEPT break down when fluid particle trajectories begin to intersect, simply because the single-stream approximation breaks down.}

In the last 20 years it has become very fruitful to apply the Eulerian and Lagrangian approach also to General Relativity (GR) \cite{Matarrese:1994wa,Matarrese:1993zf,Matarrese:1995sb,Russ:1995eu,Buchert:2012mb,Rampf:2012pu,Rampf:2013ewa,Rampf:2013dxa,Buchert:2013qma}. 
In contrast to the displacement field in the Newtonian theory, its relativistic counterpart is generally not only spatial but also contains a time-like part \cite{Ehlers:1993gf}. Indeed, GR allows for an infinite class of 4-displacements, and each displacement is associated with a given Eulerian frame \cite{Rampf:2012pu,Rampf:2013ewa,Rampf:2013dxa} (see below for a definition of Eulerian frame). 
The Eulerian frame to choose is the one in which the actual physical quantities 
(the observer wants to describe) can be most easily interpreted. So the choice of the Eulerian frame fixes the 4-displacement and vice versa, whereas the Lagrangian frame is  uniquely identified with a synchronous/comoving coordinate system \cite{Ehlers:1993gf}. The (relativistic) Eulerian frame, on the other hand, can be identified with any gauge or local coordinate system where the spatial gauge-coordinate can be identified with a Eulerian field coordinate, 
$\fett{x}(t,\fett{q})=\fett{q}+\fett{F}(t,\fett{q})$, where $\fett{x}$ is the said Eulerian field coordinate in that gauge, $\fett{q}$ the Lagrangian label (i.e., the spatial coordinate in the synchronous/comoving gauge), and $\fett{F}$ the Lagrangian displacement \cite{Rampf:2013dxa}. 
Speaking in the language of gauge transformations, we can define a specific Eulerian frame in a two-stage process. First, starting from the unique Lagrangian frame, \emph{any Eulerian frame} can be obtained with a \emph{spatial} gauge transformation which removes the longitudinal and transverse part from the spatial Lagrangian metric. Roughly speaking, the displacement field $\fett{F}$ carries this longitudinal and transverse information away from the spatial Lagrangian metric; 
as a consequence, the Newtonian part of the density and velocity agrees with the findings in NEPT.
 Second, \emph{a specific Eulerian frame} is obtained when fixing the temporal gauge condition between the Lagrangian and the Eulerian frame. 
Needless to say, this also fixes the time-component of the 4-displacement field.

Thus, relativistic Eulerian frames differ in general from each other because of the different used \emph{temporal gauge conditions}.
Consequently, in order to get the closest possible correspondence to Newtonian cosmology, it is often preferential to fix the 4-displacement field to be only of spatial nature, i.e., to use the (unique!)
Eulerian frame where the time-displacement is vanishing \cite{Rampf:2013ewa}. A very important example where a 3-displacement is preferred are cosmological $N$-body simulations. They usually require the validity of the Newtonian theory and thus also assume an overall cosmological time (i.e., no time-displacement). Certainly, we know from GR that there is no universal time, and consequently the Newtonian theory is nothing but an approximation of the complete relativistic theory. On the other hand, developing fully relativistic $N$-body simulations seems to be hopeless in the following years, so the recent folkore is rather to modify existing Newtonian simulations, and to include relativistic corrections in a qualitative way there (e.g., by demanding relativistic initial conditions as in \cite{Rampf:2012pu,Rampf:2013ewa,Rampf:2013dxa}).\footnote{See however recent attempts to conduct quasi-relativistic $N$-body simulations in the weak-field limit \cite{Adamek:2013wja}.} 

In this paper, we study the basic Eulerian frame in detail, i.e., the one where the temporal component of the 4-displacement is zero (and one obtains the Newtonian part of the Eulerian density field). To obtain the resulting 3-displacement  we shall use \emph{three different ways}---not only to clarify the connection to recent/past investigations in the literature, but also to obtain a deeper physical understanding. 
Since the temporal gauge condition is in these three approaches identical, we arrive at a unique Eulerian frame, which we call the basic Eulerian frame.
We try to keep the technical level in the main text to a minimum, and refer the interested reader to the rich appendix, especially Appendix~\ref{app:useful} where we give essential tools to extract the displacement field from a given 3-metric at arbitrary perturbative order.
In the main text, we specifically focus on the generation and evolution of secondary tensor perturbations (for the inclusion of primordial tensor perturbations, see the Appendix~\ref{app:waves}). Crucially, if the dynamical evolution of the tensor perturbations is not accounted for, spurious tensor artefacts occur in the Newtonian limit at second order. On the other hand, including the dynamical evolution of the tensors in the analysis, some second-order tensor perturbations  ($\propto a^2$)  which seem to be of ``Newtonian origin'' cancel out, and only pure gravitational waves and non-dynamical relativistic tensor perturbations  ($\propto a$)  survive. Transforming the Lagrangian metric to the Poisson gauge, which is another Eulerian frame, also this relativistic tensor contribution cancels out; in the tensor sector of the Poisson gauge, all that is left are (primary and) secondary gravitational waves (see Appendix~\ref{app:poisson}). 

This paper is organised as follows. In section~\ref{sec:grad} we introduce some useful notations and report the solutions for the Lagrangian frame. Section~\ref{sec:displacement} is devoted to the calculation of the displacement field in the three aforementioned approaches. In particular, section~\ref{sec:LEC} deals with the calculation in a local Eulerian coordinate system (i.e., it is embedded in the synchronous/comoving coordinate system), in~\ref{sec:gauge} we use a specific gauge transformation to obtain the displacement field in terms of the spatial gauge generator of that transformation, and in~\ref{sec:ADM} we describe the non-perturbative approach which relies on the Arnowitt-Deser-Misner (ADM) decomposition \cite{Arnowitt:1962hi}.  In section~\ref{sec:silent} we derive the magnetic part of the Weyl tensor to verify that our solutions are fully relativistic, and to show that it is excited by tensor perturbations and vector perturbations, where the latter is commonly referred to the frame dragging (see Refs.\,\cite{Bruni:2013mua,Rampf:2013dxa}). In section~\ref{sec:mass} we comment on recent claims about whether mass conservation in the Lagrangian frame is violated.
All former sections are restricted to an Einstein-de Sitter Universe (EdS; a matter dominated Universe with vanishing cosmological constant and no global curvature), for simplicity.
Then, in section~\ref{sec:LCDM}, we generalise our findings to a $\Lambda$CDM Universe with a primordial component of local non-Gaussianity.  We conclude in section~\ref{sec:concl}.

We wish to summarise some essential results briefly at this stage. At initial time, where the impact of the cosmological constant $\Lambda$ should have negligible impact, it is often sufficient to restrict to an EdS Universe. Then, with the use of the non-linear initial conditions~(\ref{nonlinearSeed}), we find the following 3-displacement and density contrast\footnote{As usual, $a(t)$ is the cosmological scale factor as a function of cosmic time, and $f_{\rm NL}$ denotes a primordial contribution of local non-Gaussianity. Furthermore, we denote 
$\varphi$ as the Gaussian initial potential given at some time $t_0$, $\mu_2= [(\nab \varphi)^2- \varphi_{|ab} \varphi^{|ab}]/2$ is the second-order kernel from NLPT, $C=3\nabq^{-2} \nabq^{-2} \mu_2/2 +\nabq^{-2} \Phi_{|a} \Phi^{|a}/2$ and $R_a = \nabq^{-2} ( \Phi_{|a} \nabq^2 \Phi -  \Phi_{|ab} \Phi^{|b} - 2  \partial_a\nabq^{-2} \mu_2)$ are kernels with purely relativistic origin, where $\nabq^{-2}$ is the inverse of the spatially flat Laplacian.
}
\begin{align} \label{eq:F_EdS}
  F_a (t,\fett{q}) &= \frac 3 2 at_0^2 \varphi_{|a} - \left( \frac 3 2 \right)^2 \frac 3 7 a^2t_0^4 \frac{\partial_a}{{\fett{\nabla}}^2} \mu_2 \nonumber \\
   &\quad+ \frac 3 2 at_0^2 \left( f_{\rm NL} - \frac 5 3 \right) \partial_a \varphi^2
       + 5at_0^2 \left(  C_{|a}  +R_a \right) \,,  
\end{align}
\begin{align}
 \delta(t,\fett{q}) &= - \frac 3 2 a t_0^2 \nabq^2 \varphi - 3 at_0^2 \Bigg[ \left( f_{\rm NL} - \frac 5 3 \right) \varphi \nabq^2 \varphi \Bigg. \nonumber \\
   &\qquad \qquad \qquad\; \quad+ \left( f_{\rm NL} + \frac{5}{12}  \right) \varphi_{|a} \varphi^{|a}  \Bigg. \Bigg]  \nonumber \\  
  &\quad+\left( \frac 3 2 \right)^2 a^2 t_0^4 \left[ \frac 5 7 (\nabq^2\varphi)^2 + \frac 2 7 \varphi_{|ab} \varphi^{|ab} \right] \,, 
\end{align}
where $\fett{q}$ are the Lagrangian coordinates.
For the magnetic part of the 4-Weyl tensor in a synchronous-comoving coordinate system,  H$_{\mu\nu}$, we find that only its following space-space components are non-vanishing
\begin{align} \label{eq:Hms_EdS}
   {\rm H}_{ms}(t,\fett{q}) &=   {\varepsilon}_{(m}\,^{a b} \Bigg( \frac a 2 \, \dot{\tildepi}_{bs)|a}^{\rm waves}   
      + 4 \sqrt{a} t_0 \nabq^{-2}
 \Big[ \varphi_{|as)} \nab^2 \varphi_{|b}  \Bigg. \nonumber  \\
  &\quad \hspace{3cm}- \varphi_{|ac} \varphi_{|bs)}^{|c} \Big]  \Bigg. \Bigg)  \,,
\end{align}
where the first term denotes secondary gravitational waves (i.e., induced through first-order scalar perturbations), and the round bracketed term arises through the non-linear frame-dragging.

Index notation: We use greek letters $\alpha, \beta, \ldots$ to indicate space-time indices, latin letters $i$, $j$, \ldots, to indicate any spatial coordinates, and $a$, $b$, \ldots, for spatial Lagrangian coordinates.  
We denote $q_a$ as the Lagrangian coordinate, which labels the initial position of a given fluid element. The Eulerian coordinate is $x_i$.
A komma ``${}_{,i}$'' denotes a partial differentiation w.r.t.~any spatial coordinate $x_i$, whereas a slash ``${_{|a}}$'' 
 denotes a partial differentiation w.r.t.~the Lagrangian coordinate $q_a$. Summation over repeated indices is assumed. If not otherwise stated, indices are raised and lowered with the Kronecker delta. Dots denote partial derivatives w.r.t.~cosmic time.
We set $c=1$.
Furthermore, in case of possible confusion, we label quantities with an ${\cal L}$ or ${\cal E}$ to indicate whether they are Lagrangian or Eulerian, respectively. Sometimes, for notational simplicity, we write $1/\nab^2$ instead of $\nab^{-2}= \Delta^{-1}$ for  the inverse of the spatially flat Laplacian. 
Terms decorated with the inverse Laplacian are thus understood to be formal solutions of the Laplace operator.


\section{Lagrangian frame: definitions and solutions}\label{sec:grad}

We begin with the definition of the comoving/synchronous line element which is
\be \label{synch}
  \dd s^2 = - \dd t^2 + a^2(t) \,\gamma_{a b}(t,\fett{q}) \,\dd q^a\, \dd q^b \,,
\ee
where $t$ is the proper time of the fluid element, and $a(t)$ is the cosmological scale factor, i.e., we assume cosmological perturbations on an Friedmann-Lema\^itre-Robertson-Walker (FLRW) background.
The spatial coordinate $\fett{q}$ is constant in time, hence it labels the initial position of a fluid element. This defines the Lagrangian frame.

Before deriving the Lagrangian displacement field (see the following sections), we introduce the relativistic solution in a synchronous/comoving coordinate system. 
Here we only review and not explicitly re-derive the well-known solutions for an irrotational dust model; explicit derivations can be found in e.g.~\cite{Tomita:1967,Matarrese:1997ay,Russ:1995eu,Rampf:2012pu,Rampf:2013ewa,Rampf:2013dxa}.
Here and in sections \ref{sec:displacement} and~\ref{sec:silent}  we assume the linear initial conditions. In a gradient expansion \cite{Comer:1994np}, the linear linitial conditions are equivalent to the following initial seed metric
\be \label{linearSeed}
  k_{ab}^{(1)}(\fett{q}) = \delta_{ab} \left[ 1+ \frac{10}{3} \Phi(t_0,\fett{q}) \right] \,,
\ee 
where $\Phi(t_0,\fett{q})$ is the primordial potential, here just a Gaussian random field for simplicity, given at some initial time $t_0$. For notational simplicity, we shall suppress its dependence in the following when there is no confusion. Note that we choose the above linear initial conditions to clarify the connection with the former literature (e.g., \cite{Tomita:1967,Matarrese:1997ay,Russ:1995eu}). In section~\ref{sec:LCDM}, when we include primordial 
non-Gaussianity in a $\Lambda$CDM Universe, we shall use non-linear initial conditions, which are more commonly used in recent investigations (e.g., \cite{Bartolo:2010rw,Bruni:2013qta,BruniHidalgoWands}). 

Using the linear initial conditions~(\ref{linearSeed}), it is straightforward to obtain the relativistic solution for the synchronous/comoving metric by the use of standard perturbation theory \cite{Matarrese:1997ay} or by the gradient expansion technique \cite{Rampf:2012pu}.
For example, by the use of the gradient expansion technique, one calculates the (non-)linear response of Einstein's equations by the use of the seed metric~(\ref{linearSeed}),
thus approximating the synchronous/comoving metric in an increasing number of (two) spatial gradients, order by order. 
The resulting Lagrangian solution up to second order (i.e., approximating up to four spatial gradients), for an EdS Universe, is
\begin{align} \label{sol1}
  \gamma_{ab} (t,\fett{q}) &=  \, k_{ab}^{(1)}   
  + 3a(t)t_0^2 \Bigg[ \Phi_{|ab} \left( 1-\frac{10}{3} \Phi \right) 
      - 5 \Phi_{|a} \Phi_{|b} \nonumber \Bigg. \\ 
 &\qquad\quad\qquad\quad\qquad\quad \qquad\quad\qquad+  \frac 56 \delta_{ab}  \Phi_{|c} \Phi^{|c} \Bigg. \Bigg] \nonumber \\
 &\quad   -\left( \frac{3}{2}  \right)^2 \frac 3 7 a^2(t) t_0^4 
   \Bigg[ 4 \Phi_{|ab} \nabq^2\Phi   \Bigg.  -  2\delta_{ab}\, \mu_2  \Bigg]  \nonumber \\ 
  &\quad+ \left( \frac{3}{2} \right)^2 \frac{19}{7} a^2(t) t_0^4 \,\Phi_{|ac} \Phi_{|b}^{|c} 
  + \pi_{ab} \,,
\end{align}
where we have ignored first-order vector and first-order tensor perturbations, and we have defined $\mu_2 := 1/2 \,[(\nab^2\Phi)^2-\Phi_{|cd}\Phi^{|cd})$. For an EdS Universe we have $a(t)=(t/t_0)^{2/3}$. Formally departing from the gradient expansion technique and instead insisting on the standard perturbation theory, we evaluate the time evolution of the traceless and divergenceless tensor $\pi_{ab} \equiv \pi_{ab}^{(2)}$ (which {includes} secondary gravitational waves and non-propagating tensor perturbations) by solving the wave equation
\be \label{eq:wave}
 \ddot{\pi}_{ab} +\frac 2 t \dot{\pi}_{ab} -\frac{1}{a^{2}} \nabq^{2}\pi_{ab} =   \frac{27}{14} t_0^4 \nabq^{2}{\cal S}_{ab} \,,
\ee
where we have defined the traceless and divergenceless source term (the ``tensor part'')
\begin{align} 
   {\cal S}_{ab}(\fett{q}) &=  \frac{\partial_{a} \partial_b}{\nabq^2} \mu_2  + \delta_{ab} \mu_2 - 2 \left( \Phi_{|ab} \nabq^2\Phi - \Phi_{|ac}\Phi_{|b}^{|c} \right) \,, \nonumber  \\ 
  &\qquad \bigg[ \partial^a  {\cal S}_{ab} =  {\cal S}_{a}^a = 0 \bigg] \,. 
\label{eq:Sab} 
\end{align}
The solution of the wave equation~(\ref{eq:wave}) can be derived by the use of Green's method and is found to be \cite{Matarrese:1997ay}
\be \label{sol:wave}
   \pi_{ab}(t,\fett{q}) =  - \frac{27}{14}a^2t_0^4\, {\cal S}_{ab}(\fett{q}) - 6 a t_0^2\, \nabq^{-2} {\cal S}_{ab}(\fett{q}) + \tildepi_{ab}(t,\fett{q}) \,, 
\ee
where $\tildepi_{ab} \equiv \, \tildepi_{ab}^{(2)} =\, \tildepi_{ab}^{\rm const}\!+ \tildepi_{ab}^{\rm waves}$ includes a constant term, $\tildepi_{ab}^{\rm const} \propto \nabq^{-2}\nabq^{-2}{\cal S}_{ab}$,  and another one which denotes gravitational waves, $\tildepi_{ab}^{\rm waves}$; its explicit form is not needed here, but see for example Eq.\,(4.38) in \cite{Matarrese:1997ay}. Note again that $\pi_{ab}$ contains only tensor perturbations, but only $\tildepi_{ab}$ are truly gravitational waves.

It is generally impossible to derive the wave equation~(\ref{eq:wave}) within the gradient expansion technique at any order, since the term $\nabq^2 \pi_{ab}$ (but also the source term on the RHS) is always of higher order compared to $\pi_{ab}$. Explicitly, at leading order in a spatial gradient expansion, we obtain from the tracefree part of Einstein's equations $\ddot\pi_{ab} + \frac 2 t \dot\pi_{ab} \simeq 0$. As a consequence, no gravitational waves are generated at any order in the gradient expansion, and the time evolution of generic tensor perturbations differs from the one as obtained from~(\ref{eq:wave}). Thus, the gradient expansion fails in predicting the tensor perturbations inside the horizon since it is indeed a long-wavelength approximation. 
As a necessary consequence, the gradient expansion technique and standard perturbation theory generally disagree in the tensor sector. Since we are mainly interested on scales close to the horizon, we choose to evaluate the time evolution of the tensor perturbations not with the gradient expansion technique but with standard perturbation theory. Thus, after having derived Eq.\,(\ref{sol1}), e.g., by the use of the gradient expansion, we obtain the time evolution of the tensor perturbations by plugging this solution into the $ij$ component of Einstein's equations, see the Appendix~\ref{app:waves}.

Note that the metric~(\ref{sol1}) contains intrinsic tensor perturbations \emph{even if we neglect the pure tensor perturbations} $\pi_{ab}$. 
To make this (for the second-order Newtonian terms) explicit, we can use Eq.\,(\ref{eq:Sab}) and rewrite our metric~(\ref{sol1}) to  
\begin{align} \label{sol1tensor}
   \gamma_{ab} (t,\fett{q}) &=  \, k_{ab}^{(1)}   
  + 3a(t)t_0^2 \Bigg[ \Phi_{|ab} \left( 1-\frac{10}{3} \Phi \right)  \Bigg. \nonumber \\
  &\qquad\qquad\qquad\qquad   - 5 \Phi_{|a} \Phi_{|b} +  \frac 56 \delta_{ab}  \Phi_{|c} \Phi^{|c} \Bigg. \Bigg] \nonumber \\
 &\quad  -\left( \frac{3}{2}  \right)^2 \frac 6 7 a^2(t) t_0^4 
   \Bigg[ \frac{\partial_{a}\partial_b}{\nabq^2} \mu_2 - {\cal S}_{ab}  \Bigg] \nonumber \\  
   &\quad  + \left( \frac{3}{2} \right)^2  a^2(t) t_0^4 \,\Phi_{|ac} \Phi_{|b}^{|c} 
  + \pi_{ab} \,.
\end{align}
Here, notice the occurrence of the intrinsic tensor perturbation $\propto a^2 {\cal S}_{ab}$, which originates from the second-order ``Newtonian'' terms in~(\ref{sol1}). (Of course, also the non-linear terms in the first line of~(\ref{sol1tensor}) excite tensor perturbations, which we shall derive below.)
Again, the excitation of these \emph{non-propagating tensor perturbations}
arises because of the non-linear terms in 
$\bar\gamma_{ab} \equiv \gamma_{ab} -\pi_{ab}$.
Although it is well-known that non-linear terms excite tensor perturbations \cite{Ananda:2006af,Malik:2008im}, it is less understood what happens with these non-propagating tensor perturbations when transformed to a Eulerian frame. In particular, the Newtonian limit of GR would be spoiled if there are surviving tensor perturbations.
As we shall see, the non-propagating tensor perturbations disappear in a certain Eulerian frame entirely (namely in the Poisson gauge), \emph{but} only if we solve for the gravitational waves according to the evolution equation~(\ref{eq:wave}). Since it is impossible
to derive such an evolution equation within the gradient expansion technique, 
Newtonian tensor perturbations survive and thus spoil the Newtonian limit in any frame. Thus, it is impossible to arrive at the Newtonian limit in the tensor sector within the gradient expansion technique.

We shall get more insight about tensor perturbations  by transforming the Lagrangian solution~(\ref{sol1}) to a local Eulerian coordinate system, i.e., by choosing a convenient (a triad) decomposition, see the following section.
Before doing so, we calculate the density for the metric~(\ref{sol1})
\begin{align} \label{deltaLPT}
  \delta(t,\fett{q}) &\simeq \sqrt{\frac{\det[k_{ab}^{(1)}]}{\det[\gamma_{ab}]}} -1   \nonumber \\
 &\simeq   -\frac{3}{2}at_0^2 \nabq^2 \Phi      +\left( \frac 3 2 \right)^2 a^2 t_0^4 F_2^{\cal L}(\fett{q}) \nonumber \\
 &\quad     + 3a t_0^2 \left( \frac 5 4 \Phi_{|a}\Phi^{|a} + \frac{10}{3} \Phi \nabq^2\Phi \right)
 \,,
\end{align}
where we have neglected an initial density perturbation $\delta_0$ 
(see Eq.\,(\ref{eq:deltaLag}) and the related footnote~\ref{footnote} later in the text)
\be \label{eq:F2Lag}
 F_2^{\cal L}(\fett{q}) = \left[ \frac 5 7 (\nabq^2\Phi)^2 + \frac 2 7 \Phi_{|ab} \Phi^{|ab} \right] \,.
\ee
The first term on the RHS in~(\ref{deltaLPT}) denotes the density in the Zel'dovich approximation, the term proportional to the round brackets denotes relativistic corrections which are suppressed on small scales and late times, and the square bracketed term is proportional to the second-order density field in NLPT.
Equation~(\ref{deltaLPT}) agrees with Eq.\,(4.39) in Ref.~\cite{Matarrese:1997ay}, where their growth function has to be replaced according to $\tau^2/6 \rightarrow 3/2 at_0^2$, and their $\varphi$ is our $-\Phi$.


\section{Lagrangian displacement field}\label{sec:displacement}

Although not directly apparent, the above metric~(\ref{sol1}) contains the Lagrangian solution together with its Lagrangian displacement. 
In this section, we describe three different approaches to obtain the unique 3-displacement field derived
\begin{itemize}
 \item \ref{sec:LEC}:  in a local Eulerian coordinate system;
 \item \ref{sec:gauge}: from a specific Eulerian gauge transformation;
 \item \ref{sec:ADM}: by the use of the ADM formalism.
\end{itemize}
In Fig.~\ref{fig:ADM} we show a simplistic sketch that compares the first two approaches. They are both perturbative. The third approach is the non-perturbative generalisation of the second approach.

Having three different techniques to obtain the identical (perturbative) result might seem to be superfluous. Our motivation to present them all is to demonstrate that we obtain a consistent picture of relativistic Lagrangian perturbation theory. Moreover, the following section also clarifies different approaches which were already used in the literature, and, where possible, we further develop these used techniques.

Let us briefly make an important technical comment. To obtain the displacement field (and other perturbations), we formally expand  the perturbations up to second order according to 
\be
  T =  T^{(1)}+ T^{(2)} + \ldots \,,
\ee
where $T$ denotes an arbitrary scalar, vector or tensor quantity. We thus do not approximate such quantities in a series of spatial gradients but by conventional techniques of standard perturbation theory \cite{Rampf:2012pu,Malik:2008im}.

\subsection{Perturbative displacement field in local Eulerian coordinates}\label{sec:LEC}

Here we obtain the displacement field not by performing a coordinate transformation but simply by decomposing the synchronous/comoving metric, i.e., the Lagrangian frame, in a convenient way.
We develop a Lagrangian frame theory where the synchronous-comoving metric $\fett{\gamma}$ is written  
as
 $\fett{\gamma} = G_{ij} \fett{\cal J}^i \otimes \fett{\cal J}^j$, where $G_{ij}$ is by definition \emph{not} $\delta_{ij}$. We comment on $\fett{{\cal J}}^i$ below. Our approach is fairly similar to the one of Refs.\,\cite{Buchert:2012mb,Buchert:2013qma}, however in our approach the coframe is not the only dynamical variable, because our $G_{ij}$ contains the dynamical information of the scalar and tensor part of $\fett{\gamma}$. 

The spatial metric $\gamma_{a b}$ of the synchronous slicing of Eq.\,(\ref{synch}) contains scalar, vector and tensor components, which account in total for 6 physical degrees of freedom.
We find it very convenient to decompose $\gamma_{a b}$ as
\begin{align}
  \gamma_{a b} &= G_{ij} \,{{\cal J}^{i}}_a \,{{\cal J}^j}_b \,, \label{eq:decomposeGamma} \\
  G_{ij} &= \delta_{ij} \left( 1 -2 B \right) +\chi_{ij} \,, \label{eq:decomposeGij} \\ 
  {{\cal J}^i}_a &= \delta^i_a + {F^i}_{|a} \,,
\label{eq:decompose}
\end{align}
where $B$ and $\chi_{ij}$ is a scalar and tensor perturbation, respectively, and ${F}^i$ will contain scalar and vector  perturbations. The tensor $\chi_{ij}$ is trace- and divergenceless, and if $\chi_{ij}$ is dynamical, it can be associated with gravitational waves. We comment on the scalar $B$ below.
 The Jacobian element ${{\cal J}^{i}}_a$ describes the inhomogeneous deformation of the spatial volume element, caused by the gravitational evolution of a fluid element on an FLRW background, and $F^i$ is defined as the spatial Lagrangian displacement field of the spatial coordinate transformation
\be
 \label{eq:spatial}
  \fett{x}(t,\fett{q}) = \fett{q} + \fett{F}(t,\fett{q}) \,, 
\ee
where $\fett{q}$ are the Lagrangian coordinates of the synchronous/comoving line element~(\ref{synch}), and $\fett{x}$ is the spatial field of the local Eulerian coordinate system.
We require scalar initial conditions, and it is because of that that the scalar $B$ is always non-zero in GR, i.e., it contains at least a space-dependent contribution, which is associated with some initial conditions of the scalar type, allocated from primordial physics. 
More generally, as we shall see $B$ can (and will!) contain also time-dependent contributions. 
These contributions arise from non-linearities inherent in GR, and they imply generally the loss of a universal time.

To obtain the displacement field of the metric~(\ref{sol1}) within the above decomposition, we use the expressions~(\ref{eq:decomposeGamma})--(\ref{eq:decompose}) up to second order. The relation for the 3-metric is thus up to second order
\begin{align} \label{tensor}
  \gamma_{ab} (t,\fett{q}) &\simeq \delta_{ab} \left[1-2B(t,\fett{q}) \right] + \chi_{ab} \nonumber \\ 
&\quad+2F_{(a|b)} \left[1-2B(t,\fett{q}) \right]
   + F_{(c|a)}F^{(c}{}_{|b)} \,.
\end{align}
Equating this Ansatz for the 3-metric with the solution~(\ref{sol1}), we can derive the relativistic displacement field $F_a$, the scalar $B$ as well as the tensor $\chi_{ab}$. To do so we have to decompose the above tensor equation into a scalar, solenoidal, transverse and tensor contribution. Solving these contributions separately at a given perturbative order, we obtain (1) from its divergence-less part 
\begin{widetext}

\,

\begin{figure}

\begin{minipage}[hbt]{8.5cm}

{\includegraphics[width=8.5cm]{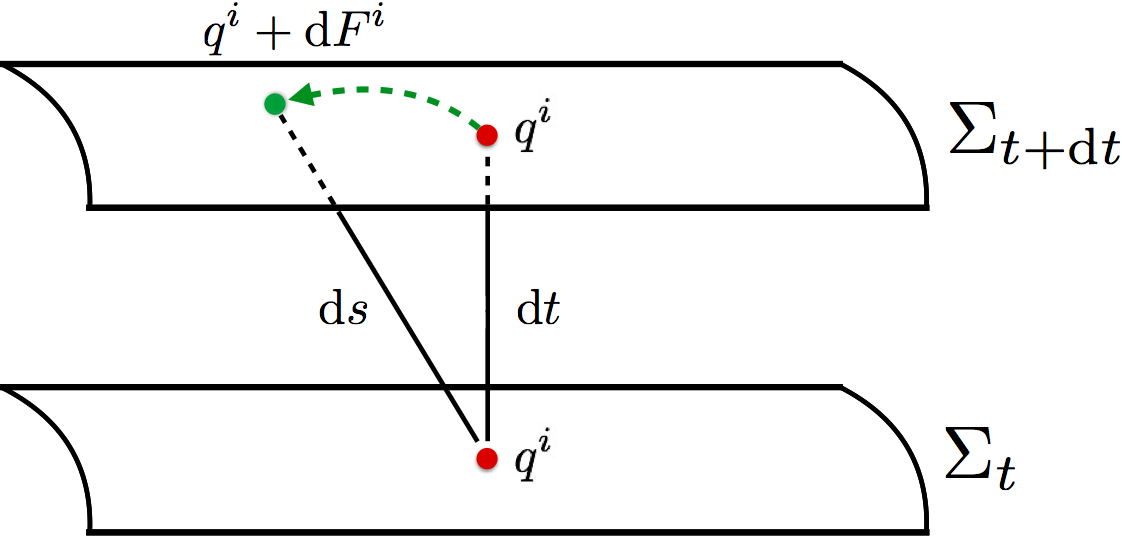}}   

\end{minipage}
\hfill
\begin{minipage}[hbt]{8.5cm}

{\vskip0.05cm\includegraphics[width=8.5cm]{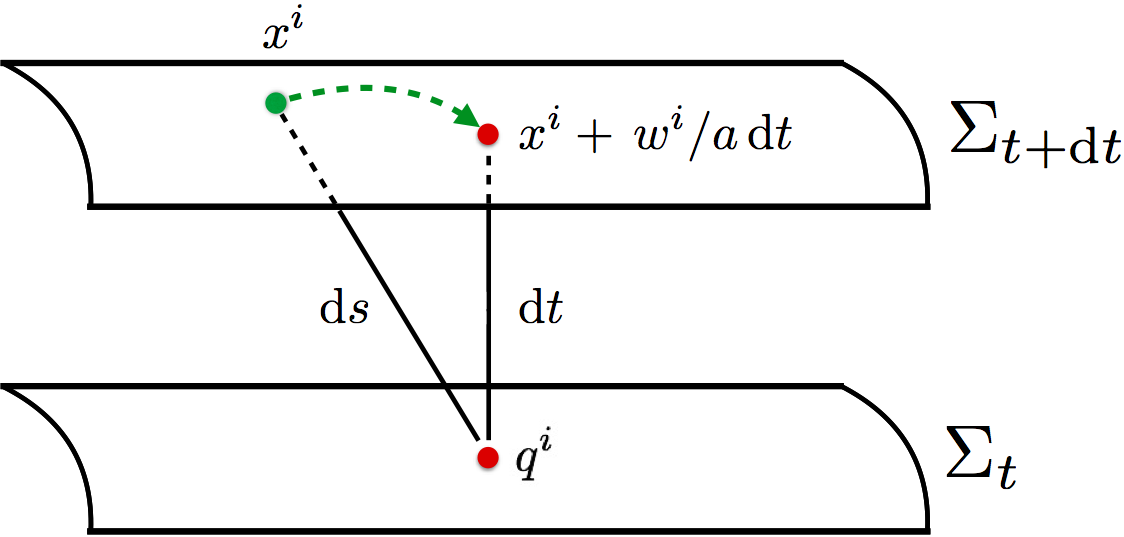}} 

\end{minipage}

\caption{Simplistic sketch of the infinitesimal 3-displacement field $\dd F^i$ (denoted as a green arrow) in the local Eulerian coordinates (LHS; see section~\ref{sec:LEC}) and in the Eulerian gauge (RHS, see section~\ref{sec:gauge}).  The final space-time positions of the fluid element (green dot) are in both cases the same.
On the LHS, we construct local Eulerian coordinates and obtain the final position of the fluid element on the space-like hypersurface $\Sigma_{t + \dd t}$ as the superposition of $q^i + \dd F^i$, where $q^i$ is the initial position of the fluid element (red dot) on $\Sigma_t$.  On the RHS, we perform a coordinate/gauge transformation \cite{Malik:2008im} from the synchronous gauge to the Eulerian gauge. Since this transformation is Lagrangian, the position $x^i$ of the fluid element on $\Sigma_{t +\dd t}$ has been already displaced/shifted by $\dd F^i = -w^i/a\,\dd t$.
The above can be also interpreted as the active (LHS) and passive (RHS) approach of a specific gauge transformation, however we prefer to see the LHS as a result of a specific tensor decomposition of the synchronous/comoving metric, as applied through Eq.\,(\ref{eq:decomposeGamma}).
\label{fig:ADM}}
\end{figure}
\end{widetext}
\noindent the transverse part of the displacement field; (2) from its solenoidal part we obtain the longitudinal part of the displacement field; (3) from its scalar part we obtain the scalar $B$. Having then derived all contributions but the tensor ones, it is simple to (4) extract the tensor parts by subtraction. We explain in Appendix~\ref{app:useful} in detail how such a decomposition works.
Here we only state the results from such a decomposition for the local Eulerian coordinates in the synchronous metric. We find
\begin{align}  \label{eq:decomp}
  \dd s^2 &= - \dd t^2 + a^2(t)\, \left[ \delta_{ij} \left( 1 -2 B \right) + \chi_{ij} \right]  \nonumber \\
    &\qquad\qquad \times\,\left[ {\delta^{i}}_a + {F^{i}}_{|a} \right] \,\left[ {\delta^{j}}_b + {F^{j}}_{|b} \right]  \,\dd q^a\, \dd q^b \,,  
\intertext{with the solutions}
  F_a(t,\fett{q}) &= \frac 3 2 at_0^2 \Phi_{|a} - \left( \frac 3 2 \right)^2 \frac 3 7 a^2t_0^4 \frac{\partial_a}{\nabq^2} \mu_2^{\cal L} \nonumber \\
     &\qquad  + 5at_0^2 \left(  C_{|a} -  \partial_a \Phi^2  +R_a \right) \,,  \label{LEC:F}\\
  B (t,\fett{q}) &=  - \frac 5 3 \Phi+  \frac 5 2 at_0^2 \left( \nabq^{-2} \mu_2^{\cal L} - \frac 1 2 \Phi_{|l} \Phi^{|l} \right) \,, \label{LEC:B} \\
 \chi_{ij} &= \,  \tildepi_{ij} - at_0^2\, \nabq^{-2} {\cal S}_{ij} \,,
 \label{LEC:chi}
\end{align}
where we have defined
\begin{align}
  C   &= \frac 3 2 \nabq^{-2} \nabq^{-2} \mu_2^{\cal L} + \frac 1 2 \nabq^{-2} \Phi_{|a} \Phi^{|a} \, , \nonumber \\
  R_a &= \nabq^{-2} \left( \Phi_{|a} \nabq^2 \Phi -  \Phi_{|ab} \Phi^{|b} - 2 \frac{\partial_a}{\nabq^2} \mu_2^{\cal L} \right) \,. \label{eq:Ra} 
\end{align}
The first term on the RHS in Eq.\,(\ref{LEC:F}) is the Lagrangian displacement field in the Zel'dovich approximation \cite{Zeldovich:1969sb}, the second term is its second-order improvement \cite{Bernardeau:2001qr}. Both terms are Newtonian and purely longitudinal. The bracketed term in Eq.\,(\ref{LEC:F}) is of purely relativistic origin, including both longitudinal and transverse contributions, respectively. The relativistic transverse contribution can be directly associated with a frame-dragging vector potential \cite{Rampf:2013dxa}.\footnote{Although not directly apparent, the kernel $C$ and the transverse vector $R_a$, Eqs.\,(\ref{eq:Ra}), are identical with Eq.\,(29) and Eq.\,(30) in \cite{Rampf:2013dxa}.}

As mentioned in section~\ref{sec:grad},
the scalar space-space perturbation $B$ contains the (linear) initial seed (the first term on the RHS in Eq.\,(\ref{LEC:B})), but also contains time-dependent terms which deform the spatial volume (``volume dilation'') of the fluid element during the gravitational evolution.  
In fact, the factor $\sqrt{1-2B}$ can be absorbed in the background scale factor $a(t) \rightarrow \tilde a(\tilde t, \fett{q})$,\footnote{This space-dependent scale factor illustrates the locally inhomogeneous expansion.} which effectively manifests
 in a temporal gauge transformation $t \rightarrow \tilde t(t,\fett{q})$; keeping aside the tensor perturbations for a moment, the resulting 3-metric is spatially flat (i.e., we arrive at the spatially flat gauge), but the time coordinate is not the one from the synchronous gauge anymore. 
  Needless to say, such a temporal gauge transformation does not change the resulting density contrast $\delta(t,\fett{q})$, if its temporal coordinate is interpreted as a function of the distorted time $\tilde t(t,\fett{q})$.

Now we comment on the tensor perturbation $\chi_{ab}$, Eq.\,(\ref{LEC:chi}). It contains two contributions, one being the gravitational waves and another tensor perturbation which is not propagating in space.  Note the partial cancellation of the non-propagating tensor perturbations in $\gamma_{ab}$ (cf.~the terms proportional to $a^2$ and $a$ in Eq.\,(\ref{sol:wave}) and Eq.\,(\ref{LEC:chi})). As mentioned above, this partial cancellation is due to the inherent tensor part in the tensor $\bar\gamma_{ab} \equiv \gamma_{ab}- \pi_{ab}$, 
i.e., in the 3-metric $\gamma_{ab}$ without the pure tensor tensor perturbation $\pi_{ab}$,
which we wish to report here explicitly
\be \label{eq:tensorpart}
  \bar\gamma_{ab}^{\rm T} \equiv  \gamma_{ab}^{\rm T}- \pi_{ab} = \frac{27}{14}a^2t_0^4\, {\cal S}_{ij} + 5at_0^2 \nabq^{-2} {\cal S}_{ij} \,, 
\ee
where $\bar\gamma_{ab}^{\rm T}$ means the transverse traceless part of $\bar\gamma_{ab}$  (so we have $\pi_{ab}^{\rm T} \equiv \pi_{ab}$ by definition). 
The crucial point about these tensor perturbations in~(\ref{eq:tensorpart}) is, that they arise through (the scalar, vector, tensor decomposition of) the non-linear terms.
Contrary to true gravitational waves, these tensor perturbations
are not propagating in space as it is the case for $\tildepi_{ab}^{\rm waves}$. Rather, these tensor perturbations are ``artefacts'', excited because of the tensorial character of Einstein's equations. In the Lagrangian frame, the first tensor perturbation on the RHS in~(\ref{eq:tensorpart}) cancels exactly out with the one from $\pi_{ab}$, whereas for the second term only the prefactor changes (cf.~Eq.\,(\ref{LEC:chi})).
 As we shall show explicitly in Appendix~\ref{app:poisson}, when we transform the Lagrangian solution~(\ref{sol1}) to the Poisson gauge, these artefacts disappear entirely such that the divergenceless and traceless part of the 3-metric in the Poisson gauge is just $\chi_{ij}^{\rm Poisson} =\, \tildepi_{ab}^{\rm waves}$.

\subsection{Perturbative displacement field from a specific gauge transformation}\label{sec:gauge}

Here we show that one can obtain the identical displacement field as above by a conceptionally different procedure, i.e.,  here we indeed perform a change of the coordinate system. This is possible if the latter coordinate system can be identified with a Eulerian frame.
We stick with the same solution $\gamma_{ab}$ as above, Eq.\,(\ref{sol1}), and transform the Lagrangian solution to the Eulerian gauge where we define the latter with  
\begin{align} \label{ADMGauge}
 \begin{split}
   \dd s^2 &= 
       - \big[ 1 \big. +2 \big. A^{\cal E}(t,\fett{x}) \big] \dd t^2  
       + 2a(t)\, w_{i}^{\cal E}(t,\fett{x}) \,\dd t\, \dd x^i  \\ 
   &\quad+ a^2(t) \,G_{ij}^{\cal E}(t,\fett{x}) \, \dd x^i \dd x^j\,,  \\
   G_{ij}^{\cal E} &=  \delta_{ij} \left[ 1- 2B^{\cal E}(t,\fett{x}) \right] + \chi_{ij}^{\cal E}(t,\fett{x})  \,, \\  
  &\qquad \qquad \left[  G_{ij}^{\perp} = G_{ij}^{\parallel} 
    = {{\chi^{i}}_{i}}^{\cal E} = \partial^i \chi_{ij}^{\cal E} = 0 \right] \,.
 \end{split}
\end{align}
and with the coordinate transformation which is in that specific case
\begin{align}  
 \label{ADMTrafo}
   x^\mu(t,\fett{q})  &= q^\mu +  F^\mu (t,\fett{q}) \,, \\  
\intertext{with}  
    x^\mu &= \begin{pmatrix} t \\ \fett{x} \end{pmatrix}, \qquad q^\mu = \begin{pmatrix} t \\ \fett{q} \end{pmatrix} \,, \qquad  F^\mu =  \begin{pmatrix} 0 \\ \fett{F} \end{pmatrix} \,. \nonumber 
\end{align}
Note that this is a purely spatial coordinate transformation, such that the time coordinate $t$ in both coordinate systems is formally identical. 
We shall see, however, that the Eulerian metric will contain a perturbation in its $\dd t^2$ component anyway (here at second order), i.e., $A\neq 0$ in general. This is however nothing but the time dilation known from special relativity due to the fluid's velocity, i.e., the $\dd t^2$ component in the Eulerian metric is the proper time.
 Also note that $\fett{w}$ contains in this gauge not only a transverse but also a longitudinal part. Roughly speaking, the above coordinate transformation shifts/pushes the 
longitudinal and transverse parts of the space-space component to the space-time component $\fett{w}$ of the metric. The Eulerian gauge has been independently introduced in Ref.~\cite{Bruni:2013qta} and Ref.~\cite{Rampf:2013dxa}; in the latter it has been (misleadingly) labelled as the synchronous-shear gauge. Also in Ref.~\cite{Rampf:2013dxa}, it has been shown that this gauge reproduces at leading order the Newtonian equations of motion.\footnote{Note that we have removed the residual gauge freedom in the above defined Eulerian gauge by explictly setting the time gauge generator in the coordinate transformation~(\ref{ADMTrafo}) to zero. Generally, the time gauge generator in the Eulerian gauge is a temporal constant $L(\fett{q})$. In reference \cite{Flender:2012nq}, the temporal displacement has been chosen ($L(\fett{q}) \neq 0$) such that the resulting Eulerian gauge yields exactly the Newtonian fluid density at leading order. Such a temporal displacement however complicates the physical interpretation. }

To obtain the displacement field and the perturbations in the Eulerian gauge, we require the invariance of the Lagrangian and Eulerian line element which reads in that case (i.e., the time coordinates are identical)
\be
  g_{\mu\nu}(t,\fett{q}) = \frac{\partial x^{\tilde \mu}}{\partial q^\mu} \frac{\partial x^{\tilde \nu}}{\partial q^\nu} g_{\tilde \mu \tilde \nu}(t,\fett{x}) \,.
\ee
Truncating up to second order, the resulting constraints between the Lagrangian metric~(\ref{sol1}) and the Eulerian metric~(\ref{ADMGauge}) are
\begin{align}
  \gamma_{ab}^{\cal L}(t,\fett{q}) &\simeq  \delta_{ab} \left[ 1 -2B^{\cal E}(t,\fett{x})\right]   +2F_{(a|b)}^{\cal L} (t,\fett{q})\left( 1-2B^{\cal E} \right)  \nonumber \\
   &\quad + F_{c|a}^{\cal L} {{F}_{|b}^{\cal L}}^{|c} + \delta_{a}^i \delta_b^j \chi_{ij}^{\cal E} \,, \label{trunc1} \\
 0 &\simeq  a^2 \left[ 1-2B^{\cal E} \right] \frac{\partial F_a^{\cal L}(t,\fett{q})}{\partial t}
  + a^2 F_{c|a}^{\cal L} \frac{\partial F^{c}_{\cal L}}{\partial t}  \nonumber \\ 
   &\quad+a \,w_a^{\cal E}(\tau,\fett{x}) + a w_{\cal E}^l F_{l|a}^{\cal L}
    \,, \label{trunc2} \\
 -1 &\simeq -1 -2A^{\cal E}(t,\fett{x})  +2a\, w_l^{\cal E} \frac{\partial F_{\cal L}^l}{\partial t}
 \label{trunc3}  +a^2 \frac{\partial F_l^{\cal L}}{\partial t} \frac{\partial F_{\cal L}^l}{\partial t} \,,
\end{align}
where we have suppressed some dependences when there is no confusion. One should evaluate the above in an identical coordinate system, e.g., $B^{\cal E}(t,\fett{x}) = B^{\cal E}(t,\fett{q}+\fett{F}) \simeq B^{\cal L}(t,\fett{q}) + B_{|a}^{\cal L} F^a + \ldots$;\footnote{When we expand the spatial dependences out, we implicitly assume small 3-displacements $\fett{F}$. That our findings from this section hold also for arbitrary large displacements follows from the non-perturbative treatment in section~\ref{sec:ADM}.} needless to say, when Eulerian [Lagrangian] quantities are derived, the Eulerian [Lagrangian] spatial dependence is needed. Note also  that all spatial derivatives in~(\ref{trunc1})--(\ref{trunc3}) are Lagrangian \emph{except} the ones inherent in $B^{\cal E}$, $w_l^{\cal E}$ and $\chi_{ij}^{\cal E}$, but such derivatives are easily transformed to Eulerian ones according to $\partial/\partial q_a = {{\cal J}^i}_a \partial/\partial_{x_i}$, when needed. 

Solving these constraints with the same techniques as above (see also \cite{Rampf:2013dxa}), we obtain for the spatial gauge generator, i.e., the 3-displacement field up to second order
\begin{align} \label{Fgauge}
  F_{a}^{\cal L}(t,\fett{q}) &= \frac 3 2 at_0^2 \Phi_{|a}(\fett{q}) - \left( \frac 3 2 \right)^2 \frac 3 7 a^2t_0^4 \frac{\partial_a}{\nabq^2} \mu_2^{\cal L} \nonumber \\
    &\quad   + 5at_0^2 \left( C_{|a} -  \partial_a \Phi^2  +R_a \right) \,.
\end{align}
This agrees exactly with our findings in section~\ref{sec:LEC}, where we calculated the displacement field in a local Eulerian coordinate system, embedded into the synchronous/comoving metric~(\ref{synch}).
For the components in the Eulerian line element~(\ref{ADMGauge}) we obtain 
\begin{align}
  A(t,\fett{x}) &= -\frac 1 2 at_0^2 \Phi_{,l} \Phi^{,l} \,,  \label{ADM_A1}\\
  B(t,\fett{x}) &= - \frac 5 3 \Phi(\fett{x}) +  \frac 5 2 at_0^2  \left[  \nabx^{-2} \mu_2^{\cal E} + \frac 1 2 \Phi_{,l} \Phi^{,l} \right]  \,,  \label{ADM_B} \\
 a\,w_i(t,\fett{x}) &= - {S}_{,i}^{\rm N} + \partial_i \left[  \frac 5 3 t \Phi^2  - \frac{10}{3} t C \right] - \frac{10}{3}t R_i  \label{ADM_w}\,, \\
  \chi_{ij}^{\cal E} &= \,\tildepi_{ij}^{\cal E} - at_0^2 \nabx^{-2} {\cal S}_{ij}^{\cal E} \,, 
\intertext{with} 
\tildepi_{ab}^{\cal L} (t,\fett{q}) &= {\cal J}^i\,_{a} {\cal J}^j\,_{b} \,\tildepi_{ij}^{\cal E}(t,\fett{x}) \,,  \nonumber 
\end{align}
where ${\cal S}_{ij}^{\cal E}$ as in Eq.\,(\ref{eq:Sab}) but with dependences and derivatives w.r.t.~the Eulerian coordinate $\fett{x}$, ${\cal J}^i\,_a$ is as defined in Eq.\,(\ref{eq:decompose}), and
\begin{align}
  \begin{split}
 {S}^{\rm N} &= \Phi(\fett{x}) \,t -  \frac 3 2 t_0^{4/3} t^{5/3} \nabx^{-2} G_2(\fett{x}) \,, \\ 
 \label{G2x} G_2 &= \frac 3 7 (\nabx^2 \Phi)^2 + \Phi_{,l} \nabx^2 \Phi^{,l} + \frac 4 7 \Phi_{,lm} \Phi^{,lm} \,.
  \end{split}
\end{align}
 $G_2$ is the well-known second-order EdS kernel for the velocity field at second order in Newtonian perturbation theory \cite{Bernardeau:2001qr}.
To calculate Eq.\,(\ref{ADM_w}), recall that the Lagrangian time derivative does not commute with the Eulerian spatial derivative. 

The interpretation of these results is as follows. Since the observer in the Eulerian frame is not comoving with the fluid element, he/she experiences a different time because of the time dilation. Actually, $-2A$ is nothing but the linear peculiar velocity squared. The scalar $B$ consists of the initial condition (here only the linear factor of $-5\Phi/3$) but also experiences a time-dependent and fully relativistic corrections (all other terms on the RHS in~(\ref{ADM_B})).
It is because of these relativistic corrections that we cannot define anymore a global time coordinate, as demanded in Newtonian physics.

${S}^{\rm N}$ is the potential  of the peculiar velocity $a\fett{u}^{\rm N}\! :=\! \nabx {S}^{\rm N}$ from Newtonian perturbation theory up to second order \cite{Rampf:2013thesis}.
In fact,  $aw_i$ (see Eq.\,(\ref{ADM_w})) contains the information about the velocity field, thus includes the relativistic contributions ${u}_i^{\rm GR}$ to ${u}_i \equiv {u}_i^{\rm N} + {u}_i^{\rm GR}$.  Additionally, as we shall see in the following section, $aw_i$ does not only contain information about the velocity field, but also contains the perturbations from the spatial metric $G_{ij}$. Specifically, we find that $w_i = - G_{ij} u^j$, with ${u}^i = a\, \partial {F}^i /\partial t$. (Again, note that we lower and raise indices with the Kronecker delta.)

From the above, it is also easy to derive the corresponding density contrast in the Eulerian gauge. It reads (cf.\,Eq.\,(\ref{deltaLPT}))
\begin{align}
 \delta(t,\fett{x}) &=  -\frac{3}{2}at_0^2 \nabx^2 \Phi  +\left( \frac 3 2 \right)^2 a^2 t_0^4 F_2(\fett{x}) \nonumber \\  
    &\quad + 3a t_0^2 \left( \frac 5 4 \Phi_{,l}\Phi^{,l} 
  + \frac{10}{3} \Phi \nabx^2\Phi \right) \,,
\end{align}
where $F_2$ is the second-order kernel from NEPT,
\be \label{eq:F2}
  F_2(\fett{x})  =\left[ \frac 5 7 (\nabx^2\Phi)^2 + \Phi_{,l}\nabx^2 \Phi^{,l} +\frac 2 7 \Phi_{,lm} \Phi^{,lm} \right] \,.
\ee
Observe the occurence of the middle term in the last expression, whereas this term is ``missing'' in the Lagrangian counterpart, c.f.~Eq.\,(\ref{eq:F2Lag}). This a well-known effect in Newtonian perturbation theory \cite{Bernardeau:2001qr}, simply stating that the Lagrangian and Eulerian mass density are fundamentally different quantities. 
Despite of recent claims \cite{Hwang:2014qfa,Yoo:2014vta}, ``the mass conservation'' is not violated.
For a discussion, see section~\ref{sec:mass}.


\subsection{Non-perturbative displacement field from the ADM decomposition}\label{sec:ADM}

In the ADM decomposition, the space-time continuum is split into spatial 
hypersurfaces $\Sigma_t$ of constant time $t$, where individual spatial hypersurfaces are seperated by the lapse function ${\cal N}$. The function ${\cal N}^i$ allows to shift within such a space-like slice.
The line element in the ADM formalism is \cite{Arnowitt:1962hi,Villa:2014aja}
\be \label{eq:adm}
  \dd s^2 = - {\cal N}^2 \dd t^2 + G_{ij} \left( a\, \dd x^i + {\cal N}^i \dd t \right)
     \left( a\, \dd x^j + {\cal N}^j \dd t \right)  \,,
\ee
with $G_{ij} := \delta_{ij}( 1 -2 B) +\chi_{ij}$ as defined in Eq.\,(\ref{eq:decompose}). 
Recall that $G_{ij}$ is generally coordinate dependent.
Since we are modelling irrotational dust, we can set immediately ${\cal N} := 1$ in the metric~(\ref{eq:adm}).
We then have
\begin{align}
  \dd s^2 &= - \dd t^2 + G_{ij} a^2 \left( \dd x^i + \frac{{\cal N}^i}{a} \dd t \right)
       \left( \dd x^j + \frac{{\cal N}^j}{a} \dd t \right)  
\label{ADMlineFirst} \\
   &= - \dd t^2 + G_{ij} a^2 \left( {{\cal J}^{i}}_a \,\dd q^a + \frac{\partial x^i}{\partial t} \dd t  + \frac{{\cal N}^i}{a} \dd t \right) \nonumber \\
     &\qquad \qquad \times  \left(  {{\cal J}^{j}}_b\, \dd q^b + \frac{\partial x^j}{\partial t} \dd t + \frac{{\cal N}^j}{a} \dd t \right) 
\label{ADMline} \,.
\end{align}
In the last line we have used the total differential of the coordinate transformation~(\ref{eq:spatial}) 
\be 
\label{totalDiff}
  \dd x^i = {{\cal J}^i}_a\, \dd q^a + \frac{\partial x^i}{\partial t} \dd t  \,.
\ee
Comparing Eq.\,(\ref{ADMline}) with Eq.\,(\ref{synch}) which we repeat here for convenience
\be
  \dd s^2 = - \dd t^2 +  G_{ij}(t,q^a) \,a \,{{\cal J}^{i}}_a \,a \,{{\cal J}^{j}}_b  \,\dd q^a\, \dd q^b \,,
\ee
we realise  that we can identify the Lagrangian frame if we set in the line element~(\ref{ADMline})
\be 
\label{eq:LagEul}
\frac{{\cal N}^i}{a} := - \frac{\partial x^i}{\partial t} \,, \qquad
  G_{ab}^{\cal L} (t,\fett{q}) = {\cal J}^i\,_{a} {\cal J}^j\,_{b} \,G_{ij}^{\cal E}(t,\fett{x}) \,.
\ee
The last expression implies $B^{\cal E}(t,{x}^i(q^a)) =  B^{\cal L}(t,{q}^a)$ and $\chi_{ab}^{\cal L} = {\cal J}^i\,_{a} {\cal J}^j\,_{b} \, \chi_{ij}^{\cal E}$. So the spatial dependences are easily transformed with the use of the spatial transformation~(\ref{eq:spatial}).
Moreover, the spatial dependences of $B^{\cal E}(t,{x}^i(q^a))$ and $B^{\cal L}(t,{q}^a)$ are dynamically related: Suppose that at initial time both Lagrangian and Eulerian frames overlap in their spatial position, i.e., $B^{\cal E}(t,{x}^i(q^a)) |_{t=t_0} =  B^{\cal L}(t_0, x^i)|_{\fett{x}=\fett{q}}$. Then the shift/displacement is just zero at initial time. After some finite time of gravitational evolution, the fluid element will have some finite coordinate velocity in the Eulerian frame (i.e., a non-vanishing shift). 
In the Lagrangian frame, since the observer is comoving with the fluid element, the shift is always zero.

Using the relations~(\ref{eq:LagEul}) in Eq.\,(\ref{ADMlineFirst}), we obtain the Eulerian description with the line element
\begin{align}
  \dd s^2 &= - \left( 1- G_{ij}^{\cal E} \left[ a \frac{\partial x^i}{\partial t} \right] \left[ a \frac{\partial x^j}{\partial t} \right] \right) \dd t^2 \nonumber \\
    &\quad  -2 G_{ij}^{\cal E}  \left[ a \frac{\partial x^i}{\partial t} \right] \dd t \,a\,\dd x^j + G_{ij}^{\cal E} \,a \,\dd x^i a \,\dd x^j \,. \label{ADMlineSecond}
\end{align}
Using the coordinate transformation~(\ref{eq:spatial}), we can even reexpress the Eulerian metric in terms of the 3-displacement field $F^i$
\begin{align}
   \dd s^2 &= - \left( 1- G_{ij}^{\cal E} \Bigg[ a \frac{\partial F_{\cal E}^i(t,\fett{x})}{\partial t} \Bigg] \left[ a \frac{\partial F_{\cal E}^j(t,\fett{x})}{\partial t} \right] \right) \dd t^2 \nonumber \\
   &\quad  -2 G_{ij}^{\cal E}  \left[ a \frac{\partial F_{\cal E}^i}{\partial t} \right] \dd t \,a\,\dd x^j + G_{ij}^{\cal E} \,a \,\dd x^i a \,\dd x^j \,, \label{ADMlineSecond2}
\end{align}
where $F_{\cal E}^i$ depends now on the Eulerian $\fett{x}$, and its Lagrangian derivatives have to be transformed to the Eulerian counterpart, e.g., for a Lagrangian derivative of any ${\cal S}(t,\fett{q})$ we have simply ${\cal S}_{|a}(t,\fett{q}) = {{\cal J}^l}_a {\cal S}_{,l}(t,\fett{x}-\fett{F})$. If we are only interested at second-order results, these transformations only matter for first-order quantities, and the dependence and derivatives of intrinsically second-order terms can just be replaced.

Note that Eq.\,(\ref{ADMlineSecond}) has been derived without any assumptions about the displacement field $F^i$; in fact, the displacement field can take arbitrary large values. Our result is fully non-perturbative, and applies for any cosmological model which assumes an FLRW background. It is also valid for general growth functions.

Relation~(\ref{ADMlineSecond2}) tells us, that if one has a Lagrangian description with a synchronous metric solution of the decomposed form~(\ref{eq:decompose}), one can immediately obtain the counterpart in the Eulerian description. 

We can also directly apply our findings to obtain fully non-perturbative relations for $A$, $B$ and $w_i$ in the case of our cosmological model from section~\ref{sec:gauge}.
Comparing the individual elements in~(\ref{ADMGauge}) and~(\ref{ADMlineSecond2}) we have
\begin{align} \label{ADM_A}
  A^{\cal E}(t,\fett{x}) &= - \frac 1 2   G_{ij}^{\cal E} \Bigg[ a \frac{\partial F_{\cal E}^i(t,\fett{x})}{\partial t} \Bigg] \left[ a \frac{\partial F_{\cal E}^j(t,\fett{x})}{\partial t} \right]  \,, \\
 B^{\cal E}(t,\fett{x}(\fett{q})) &= B^{\cal L}(t,\fett{q}) \,, \nonumber \\ 
 \chi_{ij}^{\cal L}(t,\fett{x}(\fett{q})) &= {\cal J}^i\,_{a} {\cal J}^j\,_{b}\,\chi_{ij}^{\cal E}(t,\fett{q}) \,,   \label{ADM_B_chi} \\
 w_i^{\cal E}(t,\fett{x}) &= - G_{ij}^{\cal E} \left[ a \frac{\partial F_{\cal E}^j}{\partial t} \right] \,. \label{ADM_F}
\end{align}
This is one of our main results. The square bracketed terms in the above expressions originate from the non-perturbative peculiar velocity of the fluid element, evaluated at its Eulerian position. 

It is straightforward to verify the above expressions by the use of our second-order results for $\fett{F}$ and $B^{\cal L}$ from section~\ref{sec:LEC}, see Eq.\,(\ref{LEC:F}) and Eq.\,(\ref{LEC:B}), respectively,  however one should keep in mind to transform not only the spatial dependence but also the spatial derivatives. As explained thoroughly above, this is however trivial when the solution of the displacement field is known. 

Note explicitly, that our reported results also hold with the inclusion of tensor perturbations. Whenever tensor perturbations occur at a given order, the tensor perturbations at the higher order will contain spurious elements coming from the spatial coordinate transformation, dictated by relation~(\ref{ADM_B_chi}).  (In our case, since the tensor perturbations are second order, this will influence the third-order tensors and beyond.) 
Thus, tensor perturbations are not gauge invariant beyond leading order \cite{Matarrese:1997ay,Malik:2008im}.

\section{Magnetic part of the Weyl tensor}\label{sec:silent}

Having derived the above relations, it is reasonable to ask whether our calculations are fully relativistic. Here we show that our results do indeed excite the magnetic part of the Weyl tensor, as long as tensor perturbations are included in the analysis. For convenience, we shall restrict the following derivations to the Lagrangian frame.

The Weyl tensor ${\rm C}_{\alpha \beta \gamma \delta}$ is defined to be the trace-less part of the Riemann tensor
\begin{align}
  {\rm C}_{\mu\nu\kappa\lambda} &= {\rm R}_{\mu\nu\kappa\lambda} - \left( g_{\mu[\kappa} g_{\nu\sigma]} {{\rm R}^\sigma}_\lambda + g_{\mu[\sigma}g_{\sigma\lambda]} {{\rm R}^\sigma}_\kappa \right) \\ \nonumber 
  &\quad+ \frac{\rm R}{3} g_{\mu[\kappa} g_{\nu\lambda]}  \,.
\end{align}
We define the magnetic part of the Weyl tensor as \cite{Bertschinger:1994nc}
\be 
 {\rm H}_{\mu\nu} = \frac{\sqrt{-g}}{2}u^{\kappa}u^{\lambda}\varepsilon_{\alpha\beta\kappa(\mu} {{\rm C}^{a\beta}}_{\nu)\lambda} \,,
\ee
where $u^\mu$ is the 4-velocity, $\varepsilon_{\alpha_{1}(N) \alpha_{2}(N) \cdots \alpha_{N}(N)}$ is the $N$-dimensional Levi-Civita symbol, with $\alpha_i(N) = 0,1,\ldots,N$, and $g = \det[g_{\mu\nu}]$. Here, the $g_{\mu\nu}$'s are the second-order metric coefficients in the Lagrangian frame, see Eq.~(\ref{sol1}). In the following, we shall report our results for the decomposition~(\ref{eq:decomp})--(\ref{LEC:chi}), but we have also checked that we arrive at the same result if we had instead chosen Eq.~(\ref{sol1}) as the starting point.

 In the Lagrangian frame, we have $u^0=1$ and $u^i=0$, so the above expression for ${\rm H}_{\mu\nu}$ simplifies dramatically, and we find the only non-vanishing (spatial) components
\be
 \label{magWeyl}
  {\rm H}_{ms}(t,\fett{q}) = \frac{\sqrt{-g}}{2} \varepsilon_{ab0(m} {\rm C}^{ab}\,_{s)0} \,,
\ee
where the ``$0$'' indicates a time derivative with respect to $t$.  Plugging in the decomposed form of the synchronous metric (cf.~Eq.\,(\ref{eq:decomp})), and truncate the expressions up to second order, we first obtain
\begin{align} \label{eq:HMSmain1}
  \,{\rm H}_{ms} &= \frac{\sqrt{-g}}{2 a^{2}} \Bigg\{ {\varepsilon}_{(m}\,^{a b} \left( {\dot \chi}_{bs)|a}^{\,} -2  B_{|a}\,  \dot F_{b|s)} \right)  \Bigg. \nonumber \\
  &\qquad \quad \qquad + \Bigg.  {\varepsilon}_{(m}\,^{a b} \partial_{s)}  \left[ \dot F_{b|a} - F_{a|c} \dot F_{b}^{|c} \right]  \Bigg\} \,,
\end{align}
where $\fett{F}$,  $B$ and $\chi_{bs}$ can be found in Eqs.\,(\ref{LEC:F})--(\ref{LEC:chi}). On the RHS of the above expression, the square-bracketed terms combined with the Levi-Civita symbol are called the Cauchy invariants, and state that the ``Newtonian part'' of the fluid velocity is irrotational \cite{Rampf:2012up,Zheligovsky:2013eca}. Thus, the very term is only non-zero for (anti-symmetric) non-Newtonian contributions (or if the fluid velocity is rotational); the square bracketed term yields a term 
$\propto R_a$ which can be interpreted as a source of the frame-dragging (for the definition of $R_a$, see Eq.\,(\ref{eq:Ra})). Also, the round bracketed term in~(\ref{eq:HMSmain1}) consists of non-Newtonian contributions only, so we are left with the purely relativistic expression
\begin{align}
 {\rm H}_{ms} &= \frac{\sqrt{-g}}{2 a^{2}}  {\varepsilon}_{(m}\,^{a b}
\Bigg(  {\dot \chi}_{bs)|a}^{\,} \Bigg.  \nonumber \\ 
&\quad+  \Bigg. 5 H a t_0^2  \left[ \frac{\partial_c \partial^c}{\nab^2} \Phi_{|a} \Phi_{|bs)} + \partial_{s)} R_{b|a} \right] \Bigg) \,,
\end{align}
where we have expanded the last term with a $\nab^2 \Delta^{-1} \dot = 1$, and  $R_a$ is given in Eq.\,(\ref{eq:Ra}). Since the above is already second order, we can approximate $\sqrt{-g}\simeq a^3$. Then, our final result up to second-order is
\begin{align} \label{HMSfinal}
  {\rm H}_{ms} &= {\varepsilon}_{(m}\,^{a b} \Bigg( \frac a 2 \, \dot \tildepi_{bs)|a}^{\rm waves}  \Bigg. \nonumber \\
  &\quad + \Bigg. 4 \sqrt{a} t_0 \nabq^{-2}
 \left[ \varphi_{|as)} \nab^2 \varphi_{|b} - \varphi_{|ac} \varphi_{|bs)}^{|c} \right] \Bigg)   \,.
\end{align}
This holds both for linear and non-linear initial conditions with primordial non-Gaussianity (see section~(\ref{sec:LCDM})), i.e., with the initial conditions given in~(\ref{linearSeed}) or~(\ref{nonlinearSeed}). 
As it can be seen from the above, H$_{ms}$ is excited through tensor contributions, see Eq.\,(\ref{LEC:chi}); the first originates from the gravitational waves and the second one is due to non-propagating post-Newtonian tensor perturbations.\footnote{For an EdS Universe, the \emph{Newtonian limit} of our expression~(\ref{magWeyl}) has been recently given in Ref.~\cite{Buchert:2012mb}, see their Eq.\,(83). Unfortunately, their expression contains a typo. In their notation, their Eq.\,(83) should read  $H_{j}^{i}=-\frac{1}{2J}\delta_{ab}\epsilon^{ikl}f_{|k}^{a}\left(\dot{f}_{|jl}^{b}-\dot{f}_{|m}^{b}h_{,c}^{m}f_{|jl}^{c}\right)$.}

References \cite{Matarrese:1994wa,Matarrese:1993zf} have also calculated H$_{ms}$ in the synchronous/comoving gauge, however their results do not agree with ours. We speculate that our results disagree, because the authors have not solved the wave equation~(\ref{eq:wave}) by the use of Green's method but approximately solved the wave equation in two separate regimes, i.e., inside and outside of the horizon. As a consequence of this approximative treatment in \cite{Matarrese:1994wa,Matarrese:1993zf}, gravitational waves are discarded and tensor ``artefacts''  are  excited (i.e., tensor perturbations which are non-propagating and generally of Newtonian and post-Newtonian origin, see the related discussion in section~\ref{sec:LEC}). In fact, the gravitational waves they claim to have found are not gravitational waves but indeed non-propagating tensor perturbations of Newtonian origin.
Thus, despite the fact that the authors have used standard perturbation theory to arrive at their ${\rm H}_{ms}$, they obtain an approximative result on large scales which corresponds to the long-wavelength approximation (see the related discussion after Eq.\,(\ref{sol:wave})).

\section{Is mass conservation violated in the Lagrangian frame?}\label{sec:mass}

Here we wish to comment on recent claims that the mass conservation in the Lagrangian frame is violated \cite{Hwang:2014qfa,Yoo:2014vta}. The authors give essentially two arguments for that claim. 

\subsection{The missing ``dipole term'' in the density} First, the authors of \cite{Hwang:2014qfa,Yoo:2014vta} claim that in the synchronous/comoving gauge (the ``$B$-gauge'' in \cite{Yoo:2014vta}), mass conservation is violated at second order because the density contrast does not have a ``dipole term''.
From subtracting the second-order kernels for the density, see our Eqs.\,(\ref{eq:F2Lag}) and~(\ref{eq:F2}),
 we can extract that dipole term $F_2^{\cal E} - F_2^{\cal L} = \Phi_{,l}\nabx^2 \Phi^{,l}$. This term is however well understood as it is not tied to a general relativistic description, but already appears at the Newtonian level \cite{Bernardeau:2001qr}; the dipole term appears when relating the Lagrangian mass density, $\delta^{\cal L}(t,\fett{q})$, to the Eulerian mass density, $\delta^{\cal E}(t,\fett{x})$. Then, expanding the explicit coordinate dependence up to second order, we have 
\begin{align}
   \delta^{\cal E}(t,\fett{x}) &= \delta^{\cal E}(t,\fett{q}+\fett{F}) 
     = \delta^{\cal L}(t,\fett{q}) + F^{a}\delta_{|a}^{\cal L} + \ldots \\
  &\qquad\left[\fett{x}(t,\fett{q})=\fett{q}+\fett{F}(t,\fett{q}) \right] \,, \nonumber 
\end{align}
where the second-order part of the last combination on the RHS is, apart from a time coefficient, the said dipole term.
Physically, the Lagrangian and Eulerian mass densities are fundamentally different 
quantities. The Lagrangian mass density describes the change of (mass per) volume of a given fluid element with Lagrangian label $\fett{q}$. The Eulerian mass density, however, is a field that describes the change of \emph{all} Lagrangian fluid elements. Our arguments are not tied to a given perturbative description, neither it is to cosmology in specific, but rather are a concept of general fluid mechanics. 

Related to the above, the authors of \cite{Yoo:2014vta} state that the (Newtonian) 
second order density contrast in the Lagrangian frame does not vanish on average, while the one in the Eulerian frame does. 
This argument is however flawed by the fact that they calculate the very averages of the Lagrangian and Eulerian density in a \emph{hypothetically identical} Fourier space, and this is by construction wrong. 
Explicitly, the non-local relations between the Lagrangian and Eulerian densities in their real and their Fourier spaces is
\begin{widetext}
\begin{align}
  \begin{split}
     \delta^{\cal L}(t,\fett{q}) &= \int \frac{\dd^3 \fett{k}}{(2\pi)^3} \e^{-\ii \fett{k} \cdot \fett{q}} \tilde\delta^{\cal L}(t,\fett{k}) \qquad \qquad \leftrightarrow   \qquad \qquad 
     \tilde\delta^{\cal L}(t,\fett{k}) = \int \dd^3 \fett{q} \,\e^{\ii \fett{k} \cdot \fett{q}}  \delta^{\cal L}(t,\fett{q})
\\
   &\qquad  \qquad  \updownarrow \qquad \qquad \qquad \qquad \qquad \qquad \qquad \qquad   \qquad \qquad \qquad \updownarrow \\  
     \delta^{\cal E}(t,\fett{x}) &= \int \frac{\dd^3 \fett{k}}{(2\pi)^3} \e^{-\ii \fett{k} \cdot \fett{x}} \tilde\delta^{\cal E}(t,\fett{k})  \qquad \qquad \leftrightarrow   \qquad \qquad \tilde\delta^{\cal E}(t,\fett{k}) = \int \dd^3 \fett{x} \,\e^{\ii \fett{k} \cdot \fett{x}}  \delta^{\cal E}(t,\fett{x})  \,.
  \end{split}
\end{align}
\end{widetext}
Crucially, the Fourier conjugated variable of the Eulerian [Lagrangian] Fourier space is the Eulerian  $\fett{x}$ [Lagrangian $\fett{q}$], and  the Fourier spaces are related via the displacement field in a non-linear fashion. Indeed, when properly taking the different Fourier spaces into account, also the Lagrangian density leads to a vanishing average, as it should (see, e.g., \cite{Rampf:2012xa,Matsubara:2007wj,Rampf:2012xb}).
In conclusion, the vanishing of the dipole term in the Lagrangian picture does not violate ``the'' mass conservation,  because the Eulerian and Lagrangians densities are physically not equivalent, so there is no unique mass conservation to satisfy. Of course, there is a common mutual agreement on the fact that one should be careful whether the Lagrangian and Eulerian quantity is needed in a given scenario, e.g., in case of calculating matter polyspectra we have to stick with the Eulerian mass density field \cite{Hwang:2014qfa,Yoo:2014vta}.

\subsection{On gauge artefacts in the synchronous/comoving coordinates} Second, the authors of \cite{Yoo:2014vta} state that the relativistic perturbation scheme in the synchronous comoving coordinates does not resemble the Lagrangian perturbation theory. Their argument is that there is an integration constant which does not have any counterpart in (Newtonian) LPT. In our language, their statement can be rephrased to the relation of the spatial coordinates
\be
   \fett{x}(t,\fett{q}) = \fett{q} + \fett{F}(t,\fett{q}) + \fett{c}(\fett{q}) \,,
\ee
where $\fett{c}(\fett{q})$ is the said scale-dependent integration constant, and
 $\fett{x}$ are the Eulerian coordinates and $\fett{q}$
the Lagrangian spatial labels. In principle, such an integration constant does not only arise in the general relativistic description, but also in Newtonian Lagrangian perturbation theory \cite{Rampf:2012xa}; it simply states that the Lagrangian and Eulerian frame are initially not coinciding in their (spatial) positions, and this displacement induces an initial density perturbation. In fact, in this \emph{paper} we deliberately neglected this initial density perturbation $\delta_0$, since one could in principle demand initial conditions for the general growth functions $D(t)$, $E(t)$, etc. such that $\delta_0 =0$ is initially guaranteed (for the inclusion of $\delta_0 \neq 0$, see \cite{Rampf:2013ewa}).\footnote{In GR, one can however \emph{not} neglect the initial metric perturbation $k_{ij}$ in the density, since this would imply that our space-time would be flat at any time.}

Besides of the two shortcomings reported here, the authors of \cite{Hwang:2014qfa,Yoo:2014vta} neglect the inherent non-linear constraints from GR, which become important on large scales.
We refer the interested reader to the Appendix~\ref{app:horizon}, where we derive these non-linear constraints for the density field in a leading-order approximation (in spatial gradients).

\section{Results for \texorpdfstring{$\Lambda$}{L}CDM with primordial non-Gaussianity}\label{sec:LCDM}

Here we generalise our findings from above to the case of a $\Lambda$CDM Universe with primordial non-Gaussianity. We begin with the Lagrangian frame. Since the calculational steps and the physical interpretations are exactly the same as  before, we only point out the subtleties in the initial conditions used in this section, and then state the final results.

\subsection{Lagrangian frame}
 
We take for the primordial potential
\be
 \Phi = \varphi + f_{\rm NL} \varphi^2 \,,
\ee
where $\varphi$ is a Gaussian random field, and $f_{\rm NL}$ denotes a constant component of primordial non-Gaussianity (PNG).
 Additionally, to consistently include primordial non-Gaussianity, we demand non-linear initial conditions. We thus demand for the initial seed metric the non-linear expression \cite{BruniHidalgoWands,Bruni:2013qta}
\be \label{nonlinearSeed}
   k_{ab} = \delta_{ab} \exp \left\{ \frac{10}{3} \Phi \right\}  =  \delta_{ab} \left( 1 + \frac{10}{3} \Phi + \frac{50}{9} \Phi^2 + \ldots \right)  \,,
\ee
where we neglect spatial gradients which should not play any role in the long-wavelength limit. Explicitly, these initial conditions are identical with~(\ref{linearSeed}) only at the linear level.
We find up to second order \cite{Matarrese:1997ay,Bartolo:2010rw,Rampf:2013ewa}
\begin{align}  \label{eq:metricLCDM}
  \dd s^2 &= - \dd t^2 +  \gamma_{a b}(t,\fett{q}) \,a(t) \,\dd q^a\, a(t)\, \dd q^b \,, 
\intertext{with}
\label{sol2}
 \gamma_{ab} (t,\fett{q}) &=  \, k_{ab}   
  + 3 D(t)\! \Bigg[ \varphi_{|ab} \left( 1+ 2f_{\rm NL} \varphi \right)  \Bigg. \nonumber \\
  &\qquad\quad \; +  \Bigg. \left( 2 f_{\rm NL}        - \frac 5 3 \right) \varphi_{|a} \varphi_{|b} +  \frac 56 \delta_{ab}  \varphi_{|c} \varphi^{|c} \Bigg] \nonumber \\  
 &\quad
 +\left( \frac{3}{2}  \right)^2 E(t) 
   \Bigg[ 4 \varphi_{|ab} \nabq^2\varphi   \Bigg.  
  -  2\delta_{ab} \mu_2  \Bigg]  \nonumber \\
  &\quad + \left( \frac{3}{2} \right)^2 \left[ D^2(t)-4E(t)\right] \,\varphi_{|ac} \varphi_{|b}^{|c} + \pi_{ab}^{\Lambda \rm CDM} \,,
\end{align}
where $a(t)$ is now the cosmological scale factor for $\Lambda$CDM, 
and
 $\pi_{ab}^{\Lambda \rm CDM}$ is the solution of the wave equation (here we neglect linear tensor modes)
\be \label{eq:waveLCDM}
  \ddot{\pi}_{ab}^{\Lambda \rm CDM}+3H\dot{\pi}_{ab}^{\Lambda \rm CDM}-\frac{1}{a^{2}} \nabq^{2}\pi_{ab}^{\Lambda \rm CDM}=-\frac{9}{2}\frac{E}{a^{2}}\nabq^{2}{\cal S}_{ab} \,,
\ee
where $H$ is the Hubble parameter for $\Lambda$CDM.
We shall give more details about Eq.\,(\ref{eq:waveLCDM}) in Appendix~\ref{app:waves}, where we also include linear tensor modes.  For  details about the $\Lambda$CDM growth functions $D$ and $E$
we refer the reader to the same Appendix, 
but their limits in an EdS Universe for their fastest growing modes are $D \rightarrow at_0^2$ and $E \rightarrow - 3/7a^2t_0^4$. Note however that $D$ and $E$ generally contain also decaying modes which have to be fixed by appropriate initial conditions (see also Ref.~\cite{Rampf:2013ewa}).
From the solution~(\ref{sol2}) we immediately obtain the density contrast for $\Lambda$CDM in the Lagrangian frame\footnote{Although one could require that the growth functions vanish at initial time, we should nonetheless include $\delta_0$ since the double time derivatives of the growth functions can generally yield non-zero contributions to $\delta_0$, due to the equivalence principle \cite{Rampf:2013dxa}. We set it here to zero only for simplicity.   \label{footnote}
}
\begin{align} \label{eq:deltaLag}
  \delta^{\cal L}(t,\fett{q}) &= (1+\delta_0) \sqrt{\frac{\det[k_{ab}]}{\det[\gamma_{ab}]}} -1  \\
   &\!\!\!\!\,\stackrel{\delta_0=0}{\simeq} - \frac 3 2 D(t)  \nabq^2\varphi - 3 D(t) \Bigg[ \left( f_{\rm NL} - \frac 5 3 \right)\varphi \nabq^2\varphi \Bigg. \nonumber \\
  &\qquad \qquad \qquad \qquad  \Bigg. + \left( f_{\rm NL} + \frac{5}{12}  \right) \varphi_{|a} \varphi^{|a}  \Bigg]  \nonumber \\
 &\qquad + 
\left( \frac 3 2 \right)^2 \frac 1 2 \Bigg[ \!\left( D^2(t)-E(t)\right) (\nabq^2 \varphi)^2 \Bigg. \nonumber \\
 &\qquad \qquad \qquad \,\Bigg. + \left(D^2(t)+E(t)\right) \varphi_{|ab} \varphi^{|ab} \Bigg] \,, 
\end{align}
Again, this result holds for $\Lambda$CDM with primordial non-Gaussianity, and even includes decaying modes. In Refs.~\cite{Bruni:2013qta} the above has been calculated within the same cosmology, but they neglected the decaying modes. Our result agrees with their result if we restrict our expressions to the fastest growing modes.

It is also straightforward to obtain the displacement field $F_a$ and the scalar perturbation $B$ in that case. Since the derivation is exactly the same as explained in section~\ref{sec:LEC}, we just state the final result for the decomposition
\begin{align}
 \label{decompLCDM}
  \dd s^2 &= - \dd t^2 + a^2(t)\, \left[ \delta_{ij} \left( 1 -2 B \right) + \chi_{ij} \right] \nonumber \\
   &\qquad \times \,\left[ {\delta^{i}}_a + {F^{i}}_{|a} \right] \,\left[ {\delta^{j}}_b + {F^{j}}_{|b} \right]  \,\dd q^a\, \dd q^b \,,  
\intertext{which are}
 \label{LCDM_F}
  F_a^{\cal L} (t,\fett{q}) &= \frac 3 2 D \varphi_{|a} + \left( \frac 3 2 \right)^2 E \frac{\partial_a}{\nabq^2} \mu_2 \nonumber \\
   &\quad+ \frac 3 2 D \left[ f_{\rm NL} - \frac 5 3 \right] \partial_a \varphi^2
       + 5 D \left(  C_{|a}  +R_a \right) \,, \\
  B^{\cal L}(t,\fett{q}) &=  -\frac 5 3 \left[ \varphi(\fett{q}) + \left( \frac 12 + f_{\rm NL} \right) \varphi^2 \right] \nonumber \\
  &\quad +  \frac 5 2 D(t) \left( \nabq^{-2} \mu_2 - \frac 1 2 \varphi_{|a} \varphi^{|a} \right) \,, \label{LCDM_B} \\
 \chi_{ab}^{\cal L}(t,\fett{q}) &= -\frac 9 2 E(t)\, {\cal S}_{ab} + 5 D(t) \nabq^{-2} {\cal S}_{ab} +  \pi_{ab}^{\Lambda \rm CDM} \,. \label{LCDM_chi} 
\end{align}
These results are new, and they clearly show how the PNG component affects the displacement field but also the space-space scalar perturbation $B$, where the latter takes the general relativistic volume dilation into account.
The tensor perturbations are unaffected by PNG up to second order.

\subsection{Local Eulerian frame}

With our findings from section~\ref{sec:ADM} we can easily transform the quantities~(\ref{LCDM_F})--(\ref{LCDM_B}) to the Local Eulerian frame. Using the non-perturbative results~(\ref{ADM_A})--(\ref{ADM_F}), we obtain ``up to second order'' 
\begin{align}
A^{\cal E}(t,\fett{x}) &= -\frac{9}{8} {\cal H}^{2} D^2 f_D^{2}(t) \varphi_{,l}\varphi^{,l}  \,, \\
 B^{\cal E}(t,\fett{x}) &=  -\frac 5 3 \left[ \varphi(\fett{x}) + \left( \frac 12 + f_{\rm NL}\varphi^2 \right) \right]  \nonumber \\
 &\quad+  \frac 5 2 D(t) \left( \nabq^{-2} \mu_2 + \frac 1 2 \varphi_{,l} \varphi^{,l} \right)  \,, \\
  \chi_{ij}^{\cal E}(t,\fett{x}) &= -\frac 9 2 E(t)\, {\cal S}_{ij}^{\cal E} + 5 D(t) \nabx^{-2} {\cal S}_{ij}^{\cal E}   + \pi_{ij}^{\cal E} \,, \label{chiLCDM}\\
 w_i^{\cal E}(t,\fett{x}) &=  - \frac 3 2  {\cal H}  D f_D(t) \varphi_{,i} \left( 1+\frac{10}{3} \varphi \right)  \nonumber \\
 &\quad -  \left(\frac{3}{2} \right)^2   {\cal H} \frac{\partial_i}{\nabx^2}  G_2^{\Lambda{\rm CDM}}(t,\fett{x}) \nonumber \\ &\quad 
   -\frac 3 2 {\cal H} D f_D(t) \left( f_{\rm NL} - \frac 5 3 \right) \partial_i \varphi^2 \nonumber \\ 
&\quad - 5 {\cal H} D f_D(t) \left( C_{,i} +R_i \right) \,,
\end{align}
where  ${\cal H} := aH$, where $H$ is the Hubble parameter, and
\be
 G_2^{\Lambda{\rm CDM}}(t,\fett{x}) =   \left[ \frac{D^2(t)}{2} f_D(t) \nabx^2 (\varphi_{,l} \varphi^{,l})  - 2 E(t) f_E(t)  \mu_2 \right],
\ee
and we have defined the structure growth rate  $f_X(t)=  \dd \ln X / \dd\ln a$, with $X \ni \{ D,E \}$, such that $\partial_t X= H f_X X$. Note that for an EdS Universe we have $f_D \rightarrow 1$, $f_E \rightarrow 2$, $H \rightarrow 2/(3t)$ \cite{Bernardeau:2001qr}, and $G_2^{\Lambda{\rm CDM}} \rightarrow G_2 \,a^2 t_0^4$, where $G_2$ is given in Eq.\,(\ref{G2x}).

For the Eulerian density we obtain
\begin{align}
  \delta^{\cal E}(t,\fett{x}) &= - \frac 3 2 D(t)  \nabx^2\varphi(\fett{x})  +  \left( \frac 3 2 \right)^2  F_2^{\Lambda{\rm CDM}}(t,\fett{x}) \nonumber\\ 
&\quad- 3 D(t) \Bigg[ \left( f_{\rm NL} - \frac 5 3 \right)\varphi \nabx^2\varphi \Bigg.  \nonumber \\
 &\qquad \qquad \qquad + \Bigg. \left( f_{\rm NL} + \frac{5}{12}  \right) \varphi_{,l} \varphi^{,l}  \Bigg]   \,, 
\end{align}
with
\begin{align}
  F_2^{\Lambda{\rm CDM}}(t,\fett{x}) &=  \frac 1 2 \Bigg[ \!\left(D^2-E\right) (\nabx^2 \varphi)^2+
   2 D^2\, \varphi_{,l} \nabx^2\varphi^{,l}   \Bigg. \nonumber \\
&\qquad  \qquad \Bigg. + \left(D^2+E\right) \varphi_{,lm} \varphi^{,lm} \Bigg]  \,.
\end{align}
In the EdS limit the above yields $F_2^{\Lambda{\rm CDM}} \rightarrow F_2 a^2 t_0^4$, where $F_2$ is given in Eq.\,(\ref{eq:F2}). The expressions derived in this section denote our final result related with the displacement field.


\section{Conclusions}\label{sec:concl}

We have derived relativistic solutions for the Lagrangian (3-)displacement field with vanishing time-displacement for an irrotational dust component up to second order in standard perturbation theory (i.e., not the gradient expansion technique), for both an EdS and $\Lambda$CDM Universe, including a primordial component of local non-Gaussianity. Generally, GR allows for four-dimensional displacements, so the three approaches we consider here are the only ones that deliver a purely spatial Lagrangian displacement. Needless to say, in all three approaches we arrive at the same 3-displacement, in which the bulk part is governed by Newtonian Lagrangian perturbation theory, but the derived 3-displacement contains also some relativistic corrections which should become important on large scales. Since the temporal gauge condition of the four-displacement is in these three approaches identical (i.e., we demand a vanishing time-displacement), we also arrive at a physically equivalent Eulerian frame, which we call the fundamental Eulerian frame.

Our starting point is in all three approaches the relativistic solution in a synchronous/comoving coordinate system, which we identify to be the Lagrangian frame.
The corresponding synchronous/comoving metric can be obtained by the use of the gradient expansion technique or standard perturbation theory (SPT), but to solve for the time evolution of the tensor perturbations \emph{within} the horizon, we \emph{have} to rely on approximation techniques which deliver the gravitational wave equation (see the related discussion after Eq.\,(\ref{sol:wave})), otherwise the corresponding Newtonian limit contains spurious non-propagating tensor modes. Since the gradient expansion technique is not able to produce such a wave equation, it is also impossible to derive the Newtonian limit in the tensor sector within the gradient expansion technique.
To account for the tensor perturbations, we have solved the gravitational wave equation, Eq.\,(\ref{sol:wave}), in accordance with SPT,
and thus have formally departed from the gradient expansion technique.

In the first of the aforementioned approaches, we formally do not change the coordinate system to arrive at the displacement field, but instead split the synchronous metric in a convenient way (see the line element below). 
In the second approach,
we perform a Eulerian gauge transformation from the Lagrangian to a Eulerian frame, where
the displacement field acts as the spatial gauge generator. The third approach is based on the ADM approach, and can be viewed as the non-perturbative generalisation of the second approach. 

Having obtained a relativistic Eulerian/Lagrangian correspondence, we have also derived the accompanied Newtonian part and the relativistic corrections. To avoid confusion in the following discussion 
about the relativistic corrections, we stick to the first of the three aforementioned approaches, where the synchronous/comoving line element is split according to
\begin{align}
\dd s^2 &= - \dd t^2 + a^2(t) \, \gamma_{ab}(t,\fett{q})  \,\dd q^a\, \dd q^b \nonumber \\
       &= - \dd t^2 + a^2(t)\, \left[ \delta_{ij} \left( 1 -2 B \right) + \chi_{ij} \right]  \nonumber \\
     &\qquad \quad \times \left[ {\delta^{i}}_a + {F^{i}}_{|a} \right] \,\left[ {\delta^{j}}_b + {F^{j}}_{|b} \right]  \,\dd q^a\, \dd q^b \,.
\end{align}
The purely scalar perturbation ($B$) of the relativistic corrections leads to
a spatial volume dilation, which consequently leads to a small change of the mass density. The longitudinal part ${F}_a^\parallel$ contains the Newtonian and relativistic part of the longitudinal displacement field (${F}_a = {F}_a^\parallel + {F}_a^\perp$).
The vector part of the relativistic corrections, which results from scalar induced perturbations at second order, can be easily interpreted in the Lagrangian formalism as being a transverse component in the displacement field ${F}_a^\perp$ ; it leads to a relative frame dragging between the Lagrangian and Eulerian frame. We have also discussed in detail the secondary induced tensor perturbations at second order, which we can group into three distinct kinds, namely
\begin{enumerate}
 \item gravitational waves, 
 \item relativistic non-propagating tensor perturbations, and
 \item non-relativistic non-propagating tensor perturbations.
\end{enumerate} 
The latter actually cancel entirely out in the Eulerian frame, when we solve the gravitational wave equation~(\ref{eq:wave}) by the use of Green's method.
Since these non-relativistic non-propagating tensor perturbations vanish when we properly account for the time evolution of the tensors, we consider them to be rather tensor artefacts than true ``Newtonian tensor perturbations''. If we do not solve for the 
gravitational waves, these tensor artefacts survive and flaw the Newtonian limit at second order (the Newtonian limit should obviously not include any tensor contributions at any order).
 The tensor perturbations of the second kind  cancel only partially out in the synchronous/comoving coordinate system (and also in the Eulerian gauge). 
The [partial] cancellation of tensor perturbations of the second [third] kind  occurs, since we obtain tensor perturbations from two distinct sources, i.e., 
\begin{itemize}
 \item the synchronous/comoving 3-metric, $\gamma_{ab}$, contains tensor perturbations coming from the solution of the gravitational wave equation ($\pi_{ab}$);
 \item however and crucially, even without $\pi_{ab}$, the metric $\gamma_{ab}$ contains  intrinsic tensor perturbations excited through non-linear terms.
\end{itemize}
The latter tensor perturbations arise simply because one has to decompose $\gamma_{ab}$ into scalar, vector and tensor perturbations (in Appendix~\ref{app:useful} we describe how such a decomposition works at any order), and since the metric contains non-linear features, it therefore also excites tensor perturbations.
 We have shown that the tensor perturbations of the second kind are related with the frame dragging.
Not only to prove that, but also to show that our solutions are of relativistic origin, we have calculated the magnetic part of the Weyl tensor and find that it is sourced by the tensor perturbations of the first and of the second kind.
Using our formalism, we then showed that the non-propagating tensor perturbations of the second kind disappear when we transform our results to the Poisson gauge (see the Appendix \ref{app:poisson}).
Thus, the only tensor perturbations which survive in the Poisson gauge are related to actual gravitational waves.\footnote{Starting from a different formalism, the disappearance of these non-propagating tensor perturbations has been also noted in Ref.\,\cite{Matarrese:1997ay}.} This feature underpins the assertion that the Poisson gauge is generally a preferred Eulerian frame (see e.g.,~\cite{Rampf:2013ewa}).
However, when applied, e.g., to provide relativistic initial conditions in (Newtonian) $N$-body simulations, we believe that the 3-displacement we have derived here is more suitable than the 4-displacement field related with the Poisson gauge, simply because in the case of the Eulerian gauge we only have to displace the particles in their spatial position. Nonetheless, we have given the coordinate conditions which relate the Poisson gauge and the Eulerian gauge (see Eqs.\,(\ref{conditions}) in the Appendix~\ref{app:poisson}). 

Note that all reported GR corrections should be fairly suppressed at most scales of interest w.r.t.~the respective Newtonian contributions, but they can become important on large scales, especially when one includes the biasing and the gravitational evolution in redshift space in the analysis \cite{Camera:2014bwa}.

Recently, there has been claims that the mass conservation in the Lagrangian frame
is violated, and we have attributed the section~\ref{sec:mass} to give clarifying arguments why this claim seems to be flawed. First and foremost, there is no unique concept of mass conservation simply because the Eulerian and Lagrangian mass density are different quantities by construction. The Lagrangian mass density indicates the (mass per) volume of a given fluid element with Lagrangian label $\fett{q}$, whereas the Eulerian density is a field that describes the whole density map, i.e., it includes the mass conservation of all (Lagrangian) fluid elements. When properly taking this fact into account, e.g., transforming the Lagrangian density to the Fourier space of the conjugated Eulerian space, the mass is indeed conserved.

Finally, we wish to comment on our choice of initial conditions. To make connections with results known from the earlier literature, we have deliberately used in the beginning sections~\ref{sec:grad} and \ref{sec:displacement} \emph{linear} initial conditions equivalent to the linear seed metric $k_{ab}^{(1)}= \delta_{ab}(1+ \frac{10}{3}\Phi)$, whereas in sections~\ref{sec:silent} and~\ref{sec:LCDM} we have used the non-linear initial conditions $k_{ab}=\delta_{ab} \exp\left( \frac{10}{3} \Phi \right)$ instead for our calculations.
The use of the linear initial conditions is rather historically motivated, and one should use the non-linear initial conditions, since the latter are believed to be provided by inflation. Nonetheless, all final results reported here are given for linear as well as non-linear initial conditions.

\section*{Acknowledgements}

We thank Gerasimos Rigopoulos and Eleonora Villa for useful discussions and comments on the draft. C.R.~thanks Daniele Bertacca, Marco Bruni, Uriel Frisch, Juan Carlos Hidalgo, Roy Maartens, Sabino Matarrese, David Wands and Jaiyul Yoo for clarifying their arguments. 
C.R.~thanks the Albert Einstein Institute (Potsdam/Golm) for its hospitality, particularly Lars Andersson.
C.R.~acknowledges the support of the individual fellowship RA 2523/1-1 from the German research organisation (DFG). 
For algebraic manipulations, we have used the computer algebra system \texttt{CADABRA} \cite{Peeters:2007wn}.


\appendix

\section{Generalised Helmholtz-Hodge decomposition}\label{app:useful}

Here we give some useful relations which are needed 
to disentangle the various physical contributions in the synchronous metric.
The decomposition valid for any tensor $T_{ij}$ is useful \cite{Bertschinger:1993xt},
\begin{align} \label{eq:Tij}
 T_{ij} = \frac{\delta_{ij}}{3} \hat Q + \left( \partial_i \partial_j  -\frac{\delta_{ij}}{3} \fett{\nabla}^2  \right) \hat T^\parallel + 2 \hat T_{(i,j)}^\perp + \hat T_{ij}^{\rm T} \,,
\end{align}
where $\hat Q$ is the trace of $T_{ij}$; $\hat T_i^\perp$ is a divergence-free vector; and for the transverse traceless tensor, we have $\partial^j \hat T_{ij}^{\rm T}= {\delta^{ij} T_{ij}^{\rm T}} = 0$.
It is then straightforward to define the corresponding projection operators
\begin{subequations} 
\begin{align}
 \label{operators} 
   \hat T^\parallel &= \frac 3 2 \Delta^{-1} \Delta^{-1} \partial^i \partial^j T_{ij} 
             - \frac 1 2 \Delta^{-1} \hat Q \,, \\
 \varepsilon^{kli} \hat T_{i,l}^\perp &=  \Delta^{-1}  \varepsilon^{kli} \partial^j \partial^l T_{ij} \,,
\end{align}
{where $\varepsilon^{kli}$ is the Levi-Civita symbol, and $\Delta^{-1}$ is the inverse Laplacian. Having derived $\hat Q$, $\hat T^\parallel$, and $\hat T_{i}^\perp$, one can derive the tensor contributions by trivial subtraction}
\begin{align} 
  \hat T_{ij}^{\rm T} &=  T_{ij} - \frac{\delta_{ij}}{3} \hat Q - \left( \partial_i \partial_j  -\frac{\delta_{ij}}{3} \fett{\nabla}^2  \right) \hat T^\parallel - 2 \hat T_{(i,j)}^\perp  \,. \label{operators_c}
\end{align}
\end{subequations}
For vector fields, a similar decomposition as the one above is well-known, the so-called Helmholtz-Hodge decomposition. Its validity follows from the Helmholtz theorem of vector calculus.
That the tensor decomposition~(\ref{eq:Tij}) is valid at any order follows by generalising the Helmholtz theorem to tensor fields, so we call the above decomposition the generalised Helmholtz-Hodge decomposition.

Note that the (generalised) Helmholtz-Hodge decomposition involves the non-local operation $\Delta^{-1}$. In Newtonian theory, the treatment of $\Delta^{-1}$ is trivial when the boundary conditions are known. Certainly, $\Delta^{-1}$ can be generalised to be valid on (pseudo-)Riemannian manifolds, but the boundary conditions will not be space-periodic anymore. In this paper, we are only interested in the Newtonian limit of $\Delta^{-1}$, and leave a fully relativistic treatment of $\Delta^{-1}$ for a future investigation.

Equipped with the above operators, it is straightforward to calculate the various contributions from a given tensor {\it iteratively}, i.e., we first have to find the leading order solution for such a decomposition and then recursively include the next-to-leading-order corrections. All corrections are superimposed on the lower order perturbations, where the latter are superimposed on the FLRW background.
This means, that the lower order perturbations of some field will generally affect its higher order counterpart (e.g., the second-order solution for $T^\parallel$ is sourced by its first-order solution).

\subsection{Example: Decomposition as in section~\texorpdfstring{\ref{sec:LEC}}{LEC}}

For demonstrational purposes we apply the above definitions to the calculation of section~\ref{sec:LEC}. There, we demand up to second order that
\begin{align}
 \label{appSolve}
  \gamma_{ab} (t,\fett{q}) &= \delta_{ab} \left[1-2B(t,\fett{q}) \right] + \chi_{ab} +2F_{(a|b)} \left[1-2B(t,\fett{q}) \right] \nonumber \\
   &\quad+ F_{(c|a)}F^{(c}{}_{|b)} \,,
\end{align}
where the displacement can be decomposed into a longitudinal and transverse contribution, i.e., $F_a = F_{|a}^\parallel + F_a^\perp$, with $F_a^\perp = {\varepsilon_{a}}^{bc}A_{c|b}$.
Again, the LHS of~(\ref{appSolve}) is given by solving the relativistic constraints and equations of motion, and we want to derive the quantities from the RHS in terms of that solution. Here, we choose $\gamma_{ab}$ from~(\ref{sol1}). 

First, to find the longitudinal and transverse parts of the displacement field, we apply the operators~(\ref{operators})--(\ref{operators_c}) to the Eq.\,(\ref{appSolve}). We find
\begin{align}
 \label{dec:Fparallel} F^\parallel &= \frac 1 4 \left( 3 \Delta^{-1} \Delta^{-1} \partial^a \partial^b 
    - \delta^{ab} \Delta^{-1} \right) \nonumber  \\
 &\qquad \times \left[\gamma_{ab} + 4 F_{|ab}^\parallel B - F_{|al}^{\parallel}F^{\parallel\,|l}_{|b} \right] \,,
 \\
  F_a^\perp  &= 2 \,\Delta^{-1} \Delta^{-1} {\gamma_{[an,p]}}^{,np}  \nonumber \\
    &= \Delta^{-1} {\gamma_{an}}^{,n} - \Delta^{-1} \Delta^{-1} {\gamma_{pn,a}}^{,pn}  \,,
\intertext{which then yields iteratively~(\ref{LEC:F}). Note, that these and the following relations hold only for the used cosmolgy, where at first order ${F_a^\perp}^{(1)} = \chi_{ab}^{(1)} = 0$. This can be easily rectified, if needed.
Having the longitudinal and transverse parts of the displacement field, the trace of Eq.\,(\ref{appSolve}) yields iteratively $B$, and then it remains to derive the tensor contribution from $\gamma_{ab}$. We thus have}
  B &= \frac 1 2 -\frac 1 6 \delta^{ab} \gamma_{ab} + \frac 1 3 (1-2B) \nab^2 F^\parallel + \frac 1 6 F_{|lm}^\parallel F_{\parallel}^{|lm} \,,  \\
 \chi_{ab} &= \gamma_{ab} - \delta_{ab} \left( 1-2B \right) - 2F_{(a|b)} \left(1-2B \right)
   - F_{|al}^\parallel F^{\parallel \,|l}_{|b} \,.
\end{align}
It is worthwile to stretch out the derivation of the second-order solution of $F^{\parallel (2)}$. We have, step by step,
\begin{align}
  F^{\parallel (2)} &= \frac 1 4 \left(  3  \frac{\partial^a \partial^b}{\Delta^2}  - \frac{\delta^{ab}}{\Delta} \right)
    \Bigg[ - 20 at_0^2 \Phi_{|ab} \Phi- 15 at_0^2 \Phi_{|a} \Phi_{|b} \Bigg. \nonumber \\
    &\quad- \!\left(\frac 3 2 \right)^2 \frac{12}{7} a^2t_0^4 \Phi_{|ab} \nab^2\Phi
    + \left(\frac 3 2 \right)^2 \frac{12}{7} a^2t_0^4 \Phi_{|ac} \Phi_{|b}^{|c} \Bigg. \Bigg] \nonumber \\
 &=  \frac 1 4 \left(  3  \frac{\partial^a \partial^b}{\Delta^2}  - \frac{\delta^{ab}}{\Delta} \right)    
    \left[ 5 at_0^2 \Phi_{|a} \Phi_{|b} - 10 \partial_a \partial_b \Phi^2  \right]  \nonumber \\
   &\quad+ \left(\frac 3 2 \right)^2  a^2t_0^4 \Bigg[  \frac{6}{7} \Delta^{-1} \mu_2  \Bigg. \nonumber \\
&\qquad \;\qquad- \Bigg.\frac{9}{14} \Delta^{-1} \frac{\partial^a \partial_a}{\Delta} \left\{  (\nab^2\Phi)^2 -  (\Phi_{|lm}\Phi^{|lm}) \right\} 
  \Bigg]  \nonumber \\
 &= 5 at_0^2 C - 5 at_0^2 \Phi^2 - \left(\frac 3 2 \right)^2 \frac{3}{7} a^2t_0^4 \Delta^{-1} \mu_2 \,.
\end{align}
From the second to the third line  we have used the identity $\Delta^{-1} \partial^a \partial_a  F(t,\fett{q})= F(t,\fett{q})$ twice (for some arbitrary test function $F$), whose validity in the Newtonian limit can be proven in Fourier space, $\mu_2= 1/2 \,[(\nabq^2\Phi)^2-\Phi_{|cd}\Phi^{|cd})$, and $C= 3/2 \Delta^{-1} \Delta^{-1} \mu_2 +  1/ 2 \Delta^{-1} \Phi_{|a} \Phi^{|a}$.

\section{Gravitational wave equation}\label{app:waves}

Here we derive the gravitational wave equation in a $\Lambda$CDM Universe. We choose linear initial conditions (Eq.\,(\ref{linearSeed})) and begin with the 3-metric
\begin{align}
   g_{ab}/a^2 &=  \delta_{ab}\left( 1 +\frac{10}{3} \Phi \right) +\pi^{(1)}_{ab}+\pi^{(2)}_{ab}   \nonumber \\  
&\quad  +3 D \Bigg[ \Phi_{|ab}\!\left( \!1-\frac{10}{3} \Phi \!\right)\!- 5 \Phi_{|a} \Phi_{|b} +  \frac 5 6 \delta_{ab}  \Phi_{|c} \Phi^{|c}  \Bigg]    \nonumber\\
& \quad -\frac{9}{2}\,E \left({\cal S}_{ab}-\frac{\partial_{a}\partial_b}{\nabq^{2}}\mu_{2}\right) 
+  \frac{9}{4} D\,{}^{2}\,\Phi^{|c}\,_{|b}\Phi_{|ac} \,,
\end{align}
where we have used Eqs.\,(\ref{sol1}) and~(\ref{eq:Sab}). In the following we are interested in solutions for the tensor perturbations $\pi^{(1)}_{ab}$ and $\pi^{(2)}_{ab}$.
We therefore define the expansion tensor (which is, up to a sign, identical to the extrinsic curvature) as
\be
 \Theta_{j}^{i} = - g^{ik} \dot g_{kj} \,,
\ee
and derive the $ij$ component of the Einstein equations in the ADM approach \cite{Buchert:2012mb}
\begin{align} \label{eq:theta}
\dot{\Theta}_{j}^{i}+\Theta\Theta_{j}^{i}+\frac 1 4 (\Theta_{l}^{k}\Theta_{k}^{l}-\Theta\Theta){\delta_{j}^{i}}+R_{j}^{i}- \frac R 4 \delta_{j}^{i} - \frac \Lambda 2 \delta_{j}^{i} =0 \,,
\end{align}
where $\Theta = \Theta_{a}^{a}$, and $R = g^{ab} R_{ab}(g_{ab})$. Then, we obtain at first and second order respectively
\begin{align}
 \ddot{\pi}^{(1)i}\,_{\!j}\,+ 3 H \dot{\pi}^{(1)i}\,_{\!j}- \frac{\nabq^{2}}{a^2}\pi^{(1)i}\,_{\!j}   &= 0  \,, \\
\ddot{\pi}^{(2)i}\,_{\!j}\,+ 3 H \dot{\pi}^{(2)i}\,_{\!j}- \frac{\nabq^{2}}{a^2}\pi^{(2)i}\,_{\!j}  &= - \frac{9}{2} \frac{E}{a^2}\,\nabq^{2}{\cal S}^{i}\,_{\!j}  - \Xi^{(2) i}\,_{\!j} \,,
\end{align}
where $H = \dot a/a$, and the $\Lambda$CDM time coefficients obey the partial differential equations
\begin{align}
 \label{Friedmann}  \frac{\ddot{a}}{a}+\frac{1}{2}H^{2}-\frac{1}{2}\Lambda &= 0 \,, \\
 \label{evoD} \ddot{D}+3H\dot{D}-\frac{10}{9}\frac{1}{a^{2}} &= 0  \,, \\
  \label{evoE} \ddot{E}+3H\dot{E}+\frac{1}{2}\dot{D}^{2}+\frac{10}{9}\frac{D}{a^{2}} &=0  \,,
\end{align}
and we have defined 
\begin{widetext}
\begin{align}
  \Xi^{(2) i}\,_j := & -\,\dot{\pi}^{(1)ai}\dot{\pi}_{ja}^{(1)}  + \frac{1}{a^2} \partial^{a}\pi^{(1)bi}\!\!\,\partial_{a}\pi^{(1)}_{jb}
- \frac{1}{a^2}\partial^{a}\pi^{(1)bi}\,\partial_{b}\pi^{(1)}_{ja} 
+\frac{1}{8}\,\delta^{i}\,_{j}\dot{\pi}^{(1)ba}\dot{\pi}^{(1)}_{ab}-\frac{1}{2a^2}\,\delta^{i}\,_{j}\nabq^2\pi^{(1)cb}\,\pi^{(1)}_{bc} \nonumber \\
 &+\frac{10}{3a^2}\,\Phi\nabq^{2}\pi^{(1)i}\,_{j}  + \frac{5}{a^2}\,\Phi^{|a}\partial_{a}\pi^{(1)i}\,_{j}  -\frac{5}{3a^2}\,\Phi^{|a}\partial_{j}\pi^{(1)i}\,_{a}  - 3 \dot{D}\,\dot{\pi}^{(1)ai}\,\Phi_{|aj}+\frac{3}{2} \dot{D} \dot{\pi}^{(1)i}\,_{j} \nabq^2\Phi  -3 \dot{D} \dot{\pi}^{(1)a}\,_{j}\Phi^{|i}\,_{|a}   \nonumber\\
&-  \frac{3D}{a^{2}} \Phi^{|ia}\nabq^2\pi^{(1)}_{aj}  -\frac{10}{3a^2}\,\Phi^{|a}\,_{|j} \pi^{(1)i}\,_{a}+\frac{10}{3a^2} \nabq^2\Phi\, \pi^{(1)i}\,_{j}
-\frac{3D}{a^2}\,\partial_{j}\pi^{(1)ai}  \nabq^2 \Phi_{|a}\,+3\frac{D}{a^{2}}\Phi^{|ab}\partial_{a} \partial_b \pi^{(1)i}\,_{j} \nonumber\\
&
- \frac{3D}{a^{2}} \Phi^{|ab}\partial_{j} \partial_{a}\pi^{(1)i}\!\,_{b}+\frac{3D}{2a^2}\,\partial^{a}\pi^{(1)i}\!\,_{j} \nabq^2\Phi_{|a} +\frac{3D}{a^2}\,\partial^{i}\pi^{(1)ba} \Phi_{|jab} -\frac{3D}{4a^2}\,\delta^{i}\,_{j}\partial^{a}\pi^{(1)cb}\,\Phi_{|abc}
-\frac{3D}{2a^2}\,\delta^{i}\,_{j}\Phi^{|ab} \nabq^2\pi_{ab}^{(1)} \nonumber\\
&+\frac{3}{4}\,\delta^{i}\,_{\!j}\dot{\pi}^{(1)ab}\dot{D}\,\Phi_{|ab}-\frac{5}{3a^2}\,\delta^{i}\,_{j}\Phi^{|ab}\pi^{(1)}_{ab}-\frac{3}{8a^2}\,\delta^{i}\,_{j}\partial^{a}\pi^{(1)cb}\,\partial_{a}\pi^{(1)}_{bc} +\frac{1}{4a^2}\,\delta^{i}\,_{j}\partial^{a}\pi^{(1)cb}\,\partial_{b}\pi^{(1)}_{ac} \,.
\end{align}
\end{widetext}
When linear tensor modes are absent, we have $\Xi^{(2)}=0$. Also note that Eq.\,(\ref{Friedmann}) is a combination of the two Friedmann equations. Equations~(\ref{evoD})--(\ref{evoE}) are actually the evolution equations for the first-order and second-order time coefficients; they agree with the integrated versions in Ref.~\cite{Rampf:2013ewa}.

\section{Results in the Poisson gauge}\label{app:poisson}

In this section we revisit the Lagrangian transformation to the Poisson gauge \cite{Rampf:2013dxa}, and discuss in detail what happens with the tensor perturbations. For simplicity, we restrict here again to an EdS Universe.
We transform the solution $\gamma_{ab}$ from the Lagrangian frame, Eq.\,(\ref{sol1}),  to the Poisson gauge, where the latter is associated with a (preferred) Eulerian frame. Some of the results below have been already reported in Refs.~\cite{Matarrese:1997ay,Rampf:2013dxa}, but we wish to specifically focus  on the transformation of the tensor perturbations. Here we restrict to an EdS Universe, for simplicity.

We transform the Lagrangian solution with coordinates $(t,\fett{q})$
\begin{align}
  \dd s^2  &= g_{\mu\nu}(t,\fett{q}) \,\dd q^\mu \dd q^\nu \,
   =  - \dd t^2 +a^2(t)\gamma_{ab} (t,\fett{q}) \,\dd q^a \dd q^b \,, 
\intertext{to the Poisson gauge with coordinates $(\bar\tau, \bar{\fett{x}})$ and corresponding metric ($\tau$ is \emph{not} the conformal time)}
  \dd s^2 &=  g_{\tilde\mu\tilde\nu}(\tau,\fett{x}) \,\dd x^{\tilde\mu} \dd x^{\tilde\nu}         \nonumber \\   
  &= - \big[ 1 \big. +2 \big. \bar A \big] \dd \tau^2  + 2a \bar w_i \, \dd \tau \dd x^i  \nonumber \\ 
&\quad+ a^2 \left\{ \left[ 1- 2 \bar B) \right] \delta_{ij}  + \chi_{ij}^{\rm P} \right\} \dd x^i \dd x^j \,.
\end{align}
where $\bar A$ and $\bar B$ are scalar perturbations, $\bar w_i$ is a divergenceless vector perturbation, and $\chi_{ij}^{\rm P}$ is a divergenceless and traceless tensor perturbation. 
The two coordinate systems are related by the Lagrangian coordinate transformation
\begin{align}  
 \label{LtrafoGrad}
   &\bar x^\mu(t,\fett{q}) = q^\mu +  \bar F^\mu (t,\fett{q}) \,,
\intertext{with}  
    &\bar x^\mu = \begin{pmatrix} \bar \tau \\ \bar{\fett{x}} \end{pmatrix}, \qquad q^\mu = \begin{pmatrix} t \\ \fett{q} \end{pmatrix} \,, 
  \qquad \bar{F}^\mu =  \begin{pmatrix} \bar{L} \\ \bar{\fett{F}} \end{pmatrix} \,.
\end{align}
Note that we deliberately have chosen the coordinates $(\bar \tau,\bar{\fett{x}})$ instead of  $(t,{\fett{x}})$ to account for the coordinate system in the Poisson gauge, i.e., these coordinates are not equivalent to the one from the Eulerian gauge in section~(\ref{sec:LEC}). 
To derive  $(\bar \tau,\bar{\fett{x}})$ and the perturbations in the Poisson gauge, we require the following constraints for the space-space part, the space-time part and the time-time part of the metrics, which read respectively
\begin{align}
  \gamma_{ab}(t,\fett{q}) \simeq  &-\frac{\bar L_{|a} \bar  L_{|b}}{a^2} 
    +2 \frac{\bar L_{|(a} \bar  w_{b)}}{a}  +\delta_{ab} \Bigg[ 1 -2\bar B(\tau,\fett{x})  \Bigg. \nonumber \\
  &\qquad\qquad\qquad\quad\, \Bigg. + \frac{4 \bar L(t,\fett{q})}{3t} +\frac{2\bar L^2}{9t^2} - \frac{8 \bar  B \bar  L}{3t} \Bigg]  \nonumber \\
     & +2 \bar F_{(a|b)} (t,\fett{q})\left( 1-2 \bar B + \frac{4 \bar L}{3t} \right) 
   + \bar F_{l|a} \bar F_{\,\,\,|b}^{l} \nonumber \\ 
&+ \chi_{ab}^{\rm P}(\tau,\fett{x})  \,, \label{trunc1P}  \\
 0 \simeq &- \left(  1+2 \bar A  + \frac{\partial \bar L}{\partial t} \right)  \bar L_{|a}  \nonumber \\
      &+a^2(t)\!\left[ 1-2 \bar B+\frac{4 \bar L}{3t} \right]\!\frac{\partial \bar  F_a(t,\fett{q})}{\partial t} 
  + a^2 \bar F_{l|a} \frac{\partial \bar F^{l}}{\partial t} \nonumber \\
  &  +a(t) \,\bar w_a(\tau,\fett{x}) \left[ 1+\frac{2 \bar L}{3t} + \frac{\partial \bar L}{\partial t} \right] + a \bar w^l \bar F_{l|a}
    \,, \label{trunc2P} \\
 -1 \simeq &-1 -2 \bar A(\tau,\fett{x}) -2 \frac{\partial \bar L(t,\fett{q})}{\partial t}
  -4  \bar A \frac{\partial \bar L}{\partial t}  \nonumber \\
  &-\left( \frac{\partial \bar L}{\partial t} \right)^2
  +2a\, \bar w_l \frac{\partial \bar F^l}{\partial t}
  +a^2 \frac{\partial \bar F_l}{\partial t} \frac{\partial \bar F^l}{\partial t}  \label{trunc3P} \,.
\end{align}
Now, we can dramatically simplify the calculational steps, since we have already decomposed $\gamma_{ab}$ in the local Eulerian coordinates (see section~\ref{sec:LEC}), which we repeat here for convenience
\begin{align}
  \gamma_{ab} (t,\fett{q}) &= \delta_{ab} \left[1-2B(t,\fett{q}) \right] + \chi_{ab}(t,\fett{q}) \nonumber \\
 &\quad+2F_{(a|b)} \left[1-2B(t,\fett{q}) \right]
   + F_{c|a}F^{c}{}_{|b} \,,
\end{align}
where the results for $B$, $F_a$ and $\chi_{ab}$ are given in Eqs.\,(\ref{LEC:F})--(\ref{LEC:chi}). The solutions for the 4-displacement are then straightforward to obtain. They are
\begin{align} 
  \bar L(t,\fett{q}) &= \Phi(\fett{q}) \,t +  \frac 3 4 t^{5/3} t_0^{4/3} \Phi_{|c} \Phi^{|c} \nonumber \\
&\quad- \frac 9 7 t^{5/3} t_0^{4/3} \nabq^{-2} \mu_2   -\frac 7 6 t \Phi^2 +4t C\,,\\
  \bar F_a(t,\fett{q}) &=  F_a(t,\fett{q}) + at_0^2 \left(  C_{|a} +R_a \right)  \,,
\end{align}
and for the metric perturbations in the Poisson gauge,  we find
\begin{align}
  \bar A(\tau,\fett{x}) &\simeq \phi_{\rm N} - 4 C \,, \qquad     \bar B(\tau,\fett{x}) \simeq \phi_{\rm N} +\frac 8 3 C \,,  \nonumber \\
  \bar w_i(\tau,\fett{x})  &= -4 \tau^{1/3} t_0^{2/3}  R_i \,,  \\
 \chi_{ij}^{\rm P}(\tau,\fett{x}) &=  \chi_{ij} + at_0^2\, \nabq^{-2} {\cal S}_{ij} \equiv  \,\tildepi_{ij}  \,, 
\end{align}
with
\begin{align}
  \phi_{\rm N} (\tau,\fett{x}) &= - \Phi(\fett{x}) +\frac 3 2 a t_0^2 \nabx^{-2} \Bigg[ \frac 5 7 \Phi_{|ll} \Phi_{|mm} \Bigg. \nonumber \\
   &\qquad\qquad  + \Phi_{|l} \Phi_{|lmm} + \frac 27 \Phi_{|lm} \Phi_{|lm} \Bigg. \Bigg] \,.
\end{align}
Having derived the above, it is actually easy to understand where the additional terms in $\bar F_a$ and $\chi_{ab}^{\rm P}$ arise in comparison with $F_a$ and $\chi_{ab}$. These additional terms arise because of the first term on the RHS in Eq.\,(\ref{trunc1P}), which is (obviously not apparent in the local Eulerian coordinate system and) induced through space-time mixing
\be
  -\frac{\bar L_{|a} \bar  L_{|b}}{a^2} = - a t_0^2 \Phi_{|a} \Phi_{|b} \,.
\ee
Applying the operators to extract the solenoidal, transverse, and the traceless/divergenceless part (see section~\ref{app:useful}), which we respectively denote with the superscripts $\parallel$, $\perp$ and T, we find
\begin{align}
 \begin{split}
  \left[ - at_0^2\Phi_{|a} \Phi_{|b} \right]^\parallel &= - at_0^2 C \,, \\
  \left[ - at_0^2\Phi_{|a} \Phi_{|b} \right]^\perp    &= - at_0^2 \fett{R} \,, \\
  \left[ - at_0^2\Phi_{|a} \Phi_{|b} \right]^{\rm T}  &= - at_0^2  \nabq^{-2} {\cal S}_{ab} \,.
 \end{split} 
\end{align}
This explains the additional terms in $\bar F^\parallel$, $\bar F_a^\perp$ and in 
$\bar \chi_{ab}$, and we are able to state the relation between the Poissonian coordinates
$(\bar \tau, \bar{\fett{x}})$ and the one from the local Eulerian coordinate system, $(t,\fett{x})$, which is
\begin{align} \label{conditions}
 \begin{split}
  \bar \tau(t,\fett{q})   &= t + \bar L(t,\fett{q}) \,, \\
  \bar{\fett{x}}(t,\fett{q}) &= \fett{x}(t,\fett{q}) + a(t)\, t_0^2 \left[ \nab C(\fett{q})+ \fett{R}(\fett{q}) \right] \,.
 \end{split}
\end{align}
Note that the above relation is only valid up to second order, e.g., we have 
approximated $a(\bar\tau) \simeq a(t)$ since the scale factor is decorated with terms which are already second order.

\section{Non-linear initial constraints for the density up to two spatial gradients}\label{app:horizon}

In this appendix we are interested in the fully non-linear constraints at large scales, as predicted in GR. Here, our perturbation analysis differs from the one in the rest of this \emph{paper}, since we assume that on large scales terms decorated with spatial gradients are generally small. Explicitly, we resum all terms in powers of the primordial potential but keep only terms up to two spatial gradients. Here, we also restrict to scalar perturbations only, although similar considerations should hold for vector and tensor perturbations.

From the gradient expansion, we have up to two spatial gradients \cite{Rampf:2012pu}
\begin{align}
   \gamma_{ab} &= k_{ab} +\frac{9 D(t)}{20} \left( \hat R k_{ab} - 4\hat R_{ab} \right)  \nonumber \\
   &\equiv \exp\left\{ \frac{10}{3} \Phi \right\} \Bigg[ \delta_{ab}  \Bigg.  +\frac{9 D(t)}{20} \Big( \hat R \delta_{ab} \Big. \nonumber \\   
  &\qquad\qquad\qquad\,\,\,\;\;\; \Bigg.- 4 \exp \left\{ -\frac{10}{3} \Phi  \right\} \hat R_{ab} \Big. \Big) \Bigg]  \,,
\end{align}
where the non-linear seed is $k_{ab} = \delta_{ab} \exp\{10 \Phi/3 \}$, and 
$\hat R= k^{ab} \hat R_{ab}(k_{ab})$. 
The density contrast is then up to two spatial gradients
\begin{align}
   \delta(t,\fett{q}) &\stackrel{(\delta_0 \simeq 0)}{=} \sqrt{\frac{\det k_{ab}}{\det \gamma_{ab}}} -1  \nonumber \\
   &\equiv {\det \left[ \delta_{ab} + \frac{9 D}{20} ( \hat R \delta_{ab} - 4 \exp \{ -\frac{10}{3} \Phi \} \hat R_{ab} )  \right]}^{-\frac 12}\!\!-1 \nonumber \\
    & \simeq \frac{1}{\sqrt{1 -\frac{9 D(t)}{20} \hat R}} -1 \simeq
 \frac{9D(t)}{40} \hat R \nonumber \\ 
 &\equiv - D(t) \exp \left\{ -\frac{10}{3} \Phi \right\} \left[ \frac 3 2 \nabq^2 \Phi + \frac{5}{4} \Phi_{|l}\Phi^{|l} \right] \,.
\end{align}
 The last line is a very powerful result, since it can be used to estimate the inherent non-linearities in general relativity at any order on superhorizon scales \cite{BruniHidalgoWands}. On large scales, velocity terms $\propto \nabq\Phi$ do not survive, such that the above reduces to
\begin{align}
  \delta^{\rm super-horizon}(t,\fett{q}) &\simeq  \delta^{(1)}(t,\fett{q})
    \Bigg[ 1- \frac{10}{3} \Phi \Bigg. \nonumber \\
  &\qquad \Bigg. + \frac{50}{9} \Phi^2 - \frac{500}{81} \Phi^3 + \ldots \Bigg] \,, \\
    \delta^{(1)}(t,\fett{q}) &:= -\frac 3 2 D(t) \nabq^2 \Phi \,. \nonumber
\end{align} 
{Comparing this with the predictions from a cosmology with local-type primordial non-Gausianity with the large-scale contributions \cite{BruniHidalgoWands}}
\begin{align}
  \delta^{\rm super-horizon}(t,\fett{q})  &\simeq \delta^{(1)}(t,\fett{q}) \Big[ 1 + 2 f_{\rm NL} \Phi \Big. \nonumber \\
   &\quad \Big. +3 g_{\rm NL} \Phi^2 + 4 h_{\rm NL} \Phi^3 + \ldots \Big] \,,
\end{align}
we can easily read-off the inherent local non-linearities from GR on large scales
\begin{align}
 f_{\rm NL}^{\rm GR} = - \frac 5 3 \,, \qquad g_{\rm NL}^{\rm GR} = \frac{50}{27} \,,
   \qquad h_{\rm NL}^{\rm GR} = - \frac{125}{81} \,. 
\end{align}
This result is in agreement with the findings in Ref.\,\cite{BruniHidalgoWands}.

\end{document}